\documentclass[traditabstract, final]{aa}
\usepackage{graphicx}
\usepackage[small]{caption}
\usepackage{epsfig}
\usepackage{natbib}
\usepackage{fancyhdr}
\usepackage{amsmath}
\usepackage{amssymb}
\usepackage{makeidx}
\usepackage{titletoc}
\usepackage[]{subfigure}
\usepackage{lscape}
\usepackage{hyperref}
\usepackage{wasysym}
\usepackage{wrapfig}
\usepackage{longtable}
\usepackage{aas_macros}
\usepackage{txfonts}

\begin{document}

\title{The episodic Star Formation History of the Carina Dwarf Spheroidal Galaxy\thanks{Tables 1, 2 and 3 are only available in electronic form at the CDS via anonymous ftp to cdsarc.u-strasbg.fr (130.79.128.5) or via http://cdsweb.u-strasbg.fr/cgi-bin/qcat?J/A+A/}}
\titlerunning{The Star Formation History of the Carina dSph}

   \author{T.J.L. de Boer\inst{1,2}
          \and
          E. Tolstoy\inst{2} 
          \and
          B. Lemasle\inst{3}
          \and
          A. Saha\inst{4}
          \and
          E.W. Olszewski\inst{5}
          \and
          M. Mateo\inst{6} 
          \and
          M.J. Irwin\inst{1} 
          \and
          G. Battaglia\inst{7} 
  }

   \institute{Institute of Astronomy, University of Cambridge, Madingley Road, Cambridge CB3 0HA, UK\\
              \email{tdeboer@ast.cam.ac.uk}
             \and
             Kapteyn Astronomical Institute, University of Groningen, Postbus 800, 9700 AV Groningen, The Netherlands
             \and
             Astronomical Institute Anton Pannekoek, University of Amsterdam, Science Park 904, 1098 XH Amsterdam, The Netherlands
             \and
              National Optical Astronomy Observatory\thanks{The National Optical Astronomy Observatory is operated by AURA, Inc., under cooperative agreement with the National Science Foundation.},
              P.O. box 26732, Tucson, AZ 85726, USA
             \and
             Steward Observatory, The University of Arizona, Tucson, AZ 85721, USA
             \and
             Department of Astronomy, University of Michigan, Ann Arbor, MI 48109-1090, USA
             \and
             INAF~$-$~Osservatorio Astronomico di Bologna Via Ranzani 1, I$-$40127, Bologna, Italy
             }

   \date{Received ...; accepted ...}

\abstract{We present deep photometry of the Carina dwarf Spheroidal galaxy in the B,V filters from CTIO/MOSAIC, out to and beyond the tidal radius of r$_{ell}$$\approx$0.48 degrees. The accurately calibrated photometry is combined with spectroscopic metallicity distributions of Red Giant Branch stars to determine the detailed star formation and chemical evolution history of Carina. \\
The star formation history confirms the episodic formation history of Carina and quantifies the duration and strength of each episode in great detail, as a function radius from the centre. Two main episodes of star formation occurred at old~($>$8 Gyr) and intermediate~(2$-$8 Gyr) ages, both enriching stars starting from low metallicities~([Fe/H]$<$$-$2 dex). By dividing the SFH into two components, we determine that 60$\pm$9 percent of the total number of stars formed within the intermediate age episode. Furthermore, within the tidal radius~(0.48 degrees or 888 pc) a total mass in stars of 1.07$\pm$0.08$\times$10$^{6}$ M$_{\odot}$ was formed, giving Carina a stellar mass-to-light ratio of 1.8$\pm$0.8. \\
Combining the detailed star formation history with spectroscopic observations of RGB stars, we are able to determine the detailed age-metallicity relation of each episode and the timescale of $\alpha$-element evolution of Carina from individual stars. The oldest episode displays a tight age-metallicity relation during $\approx$6 Gyr with steadily declining $\alpha$-element abundances and a possible $\alpha$-element ``knee" visible at [Fe/H]$\approx$$-$2.5 dex. The intermediate age sequence displays a more complex age-metallicity relation starting from low metallicity and a sequence in $\alpha$-element abundances with a slope much steeper than observed in the old episode, starting from [Fe/H]=$-$1.8 dex and [Mg/Fe]$\approx$0.4 dex and declining to Mg-poor values~( [Mg/Fe]$\leq$$-$0.5 dex). This indicates clearly that both episodes of star formation formed from gas with different abundance patterns, inconsistent with simple evolution in an isolated system.}

\keywords{Galaxies: stellar content -- Galaxies: evolution -- Galaxies: dwarf -- Galaxies: Local Group -- Stars: C-M diagrams}

\maketitle

\section{Introduction}
\label{introduction}
Dwarf galaxies play an important role in the study of galaxy formation and evolution, since they are the most common type of galaxy in the $\Lambda$-CDM framework and they are believed to be among the first objects that formed in the Universe~\citep[e.g.][]{Kauffmann93}. The dwarf galaxies in the Local Group~(LG) are especially important, since their properties can be studied in more detail than those of more distant systems. However, the range of galaxy types and environmental effects that can be probed by studying LG dwarf galaxies is limited~\citep[e.g.,][and references therein]{Tolstoy09}. In particular, the effect of tidal encounters between galaxies is thought to play an important role in the evolution of dwarf galaxies in clusters of galaxies, but its effects are not easily probed in the LG. In this context, the Carina dwarf spheroidal~(dSph) galaxy is an important system, since it is the only LG galaxy that displays clear signs of a uniquely episodic formation history, possibly induced by tidal interactions with the Milky Way~(MW). Therefore, it could provide an important test case for theories that study the effects of environmental effects on the evolution of dwarf galaxies. \\
Carina is one of the smaller "classical" dSphs, with a total~(dynamical) mass of $\approx$3.4$\times$10$^{6}$ M$_{\odot}$ within an half-light radius of r$_{h}$=250$\pm$39 pc, and an absolute visual magnitude of M$_{\mathrm{V}}$=$-$9.3~\citep{Irwin95, Mateo98,Walker11,McConnachie12}.The distance to Carina has been determined using several optical and near-infrared distance indicators, including RR Lyrae stars, tip of the Red Giant Branch~(RGB) and the Horizontal Branch~(HB) level, resulting in a distance of~106$\pm$2 kpc or (m-M)$_{\mathrm{V}}$=20.13$\pm$0.04~\citep{DallOra03, Smecker-Hane94, Pietrzynski09}. Furthermore, Carina has a core radius of 0.15 degrees~(277 pc) and tidal radius of 0.48 degrees~(888 pc) as determined from King profile stellar density fitting and an estimated extinction of E(B$-$V)=0.061 as measured from dust extinction maps~\citep{Irwin95,Schlegel98}. \\    
Photometric studies of Colour-Magnitude Diagrams (CMDs) suggested initially that Carina was a purely intermediate age galaxy without old populations~\citep{Mould83}. Furthermore, Carina was found to have a narrow RGB, suggesting a small range in metallicity and a simple evolution. However, studies of variable stars quickly showed that Carina also contains RR Lyrae stars, indicating the presence of ancient~($>$10 Gyr) stellar populations~\citep{Saha86}. Later deep CMD studies revealed multiple main-sequence turnoffs~(MSTOs) in Carina and concluded that a significant fraction of stars formed more than 10 Gyr ago~\citep{Smecker-Hane96, Dolphin02, Monelli03, Small13}. 
\begin{figure}[!ht]
\centering
\includegraphics[angle=270, width=0.49\textwidth]{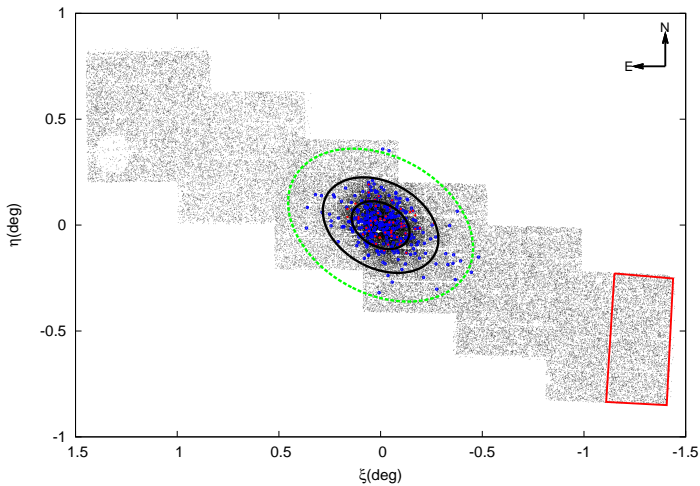}
\caption{Coverage of the photometric and spectroscopic observations across the Carina dwarf spheroidal galaxy. The CTIO 4m/MOSAIC photometry is shown as small black dots, while the open~(blue) circles show the low resolution VLT/FLAMES \ion{Ca}{ii} triplet sample and the solid~(red) dots show the high resolution spectroscopic sample. The dashed~(green) ellipse indicates the tidal radius of Carina~(28.8 arc min, 0.48 degrees), while the black ellipses indicate one and two core radii~\citep{Irwin95}. The (0,0) coordinate in the coverage plot corresponds to RA=06:41:36.70, dec=-50:57:58.00. Finally, the region outlined by the solid red line is used for MW foreground correction. The large gap in the top-left is caused by a bright foreground star which saturates the CCD. \label{Carcov}}
\end{figure}
\\
Subsequent photometric studies confirmed the presence of distinct MSTOs with a clear lack of stars in between, which is evidence of episodes of active star formation separated by periods consistent with no star formation at all~\citep[e.g.][]{Mighell97, HurleyKeller98, Hernandez00, Rizzi03, Bono10}. There are also indications of other, younger episodes as suggested by the presence of anomalous cepheids with ages $<$1 Gyr and detected in the CMD as a blue plume of young Main Sequence~(MS) stars~\citep{Mateo982, Monelli03}. There is general agreement between different studies that the intermediate age episode took place somewhere between 3 and 8 Gyr ago and the older episode $>$8 Gyr ago, but the exact age and duration of the episodes still remains uncertain. This is due in part to the different spatial area covered by different studies,  as well as the metallicity assumed for stars in the different episodes. \\
Spectroscopic studies of Carina have revealed the metallicity and abundance patterns of individual stars on the upper RGB. Medium resolution \ion{Ca}{ii} triplet spectroscopy was used to determine the metallicity [Fe/H] of Carina, reporting values between $-$1.5 and $-$2.0 dex with a small dispersion~\citep{Armandroff91, DaCosta94, Smecker-Hane99}. However, more recent studies of several hundred stars show a detailed Metallicity Distribution Function~(MDF), which displays a large range in metallicity~(0.92 dex) with a peak around [Fe/H]=$-$1.4 dex~\citep{Koch06,Helmi06}. This wide range in metallicity is much greater than expected from the narrow RGB and shows that there is a strong age-metallicity degeneracy in Carina. 
\begin{figure}[!ht]
\centering
\includegraphics[angle=0, width=0.49\textwidth]{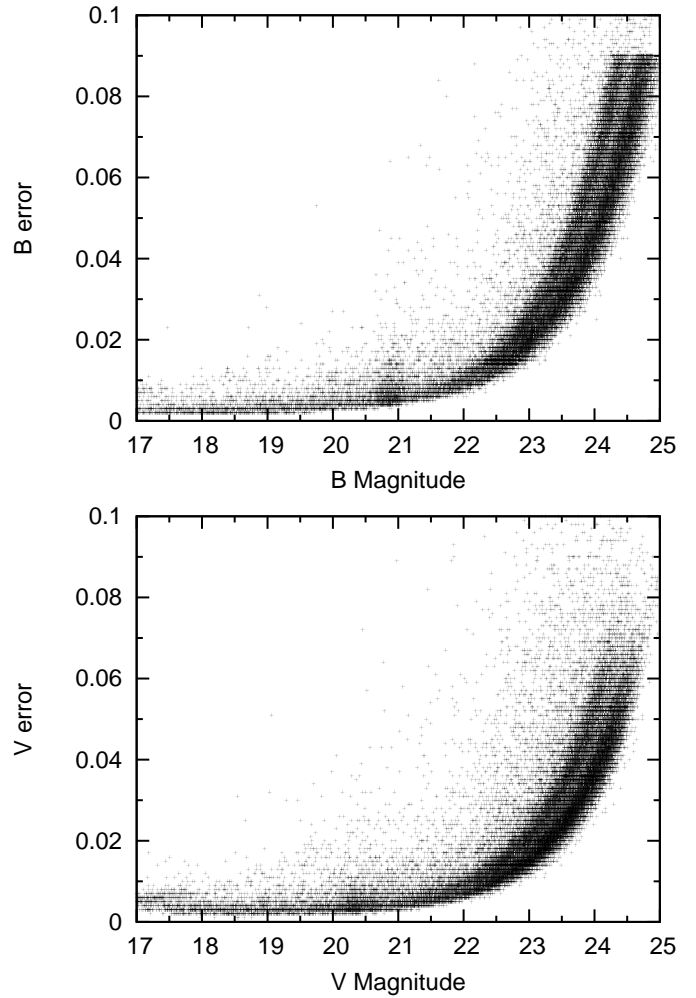}
\caption{Photometric error as a function of magnitude for the B and V filters for all stars within the tidal radius of Carina. \label{Carphoterr}}
\end{figure}
\\
High Resolution~(HR) spectroscopic studies of individual stars have revealed the complex abundance patterns in Carina, including alpha, iron-peak and heavy-elements~\citep{Shetrone03, Koch08, Lemasle12,Venn12}. These studies have revealed a significant star to star scatter in the alpha abundance ratios of Carina stars. \citet{Lemasle12} used photometric age determinations of individual stars to divide his spectroscopic sample according to age. The stars associated to the older episode display a trend of decreasing alpha-element abundances with increasing metallicity, similar to what is observed in other LG dwarf galaxies~\citep[e.g.,][]{Tolstoy04}. The stars associated to the intermediate age episode show a large scatter in alpha-element abundances for a small range in metallicities~($-$1.8$<$[Fe/H]$<$$-$1.2 dex), indicating that each star formation episode experienced a different chemical enrichment. \\
Studies of the spatial distribution of individual stars have shown that Carina displays negative age and metallicity gradients with radius~\citep{Majewski00, Harbeck01, Munoz06, Battaglia12}. This is similar to population gradients found in other LG dSphs~\citep{Tolstoy04, Battaglia06, Battaglia11, deBoer2012A, deBoer2012B}. Furthermore, a recent wide field photometric study has found signs of extra-tidal stars and possible tidal tails, suggesting that Carina likely experienced tidal encounters, which could be responsible for the episodic SFH~\citep{Battaglia12}. The occurrence of a recent tidal encounter is also consistent with proper motion determinations that place Carina in an orbit currently at apocenter with an orbital period of between 1.3 and 2 Gyr and a last pericenter passage 0.7 Gyr ago that coincides with the age of the young MS population~\citep{Piatek03}. \\
In this work, we use deep, wide field photometry obtained with MOSAIC on the CTIO 4m/Blanco telescope to determine the detailed SFH of the Carina dSph. The wide field photometry allows us to determine the SFH at different distance from the centre of Carina, to confirm and quantify the population gradients. Spectroscopic \ion{Ca}{ii} triplet metallicities are directly used in combination with the photometry, to provide additional constraints on the age of the stellar populations, as described in~\citet{deBoer2012A}. In this way, no assumptions need to be made about the metallicity for stars in the different episodes. Furthermore, including spectroscopic metallicities increases the age resolution of the recovered SFH, allowing us to resolve the star formation episodes in greater detail than ever before. \\
The paper is structured as follows: in section~\ref{data} we present the new observations used to determine the SFH of Carina. Section~\ref{SFHmethod} describes the determination of the SFH and discusses the specifics of fitting Carina. The detailed, spatially resolved SFH of Carina is presented in section~\ref{CarSFH} and the chemical evolution timescales derived from individual RGB stars in section~\ref{Carindivages}. Finally, section~\ref{conclusions} discusses the results obtained from the SFH and chemical evolution timescales.

\section{Data}
\label{data}
\begin{figure*}[t]
\centering
\includegraphics[angle=0, width=0.95\textwidth]{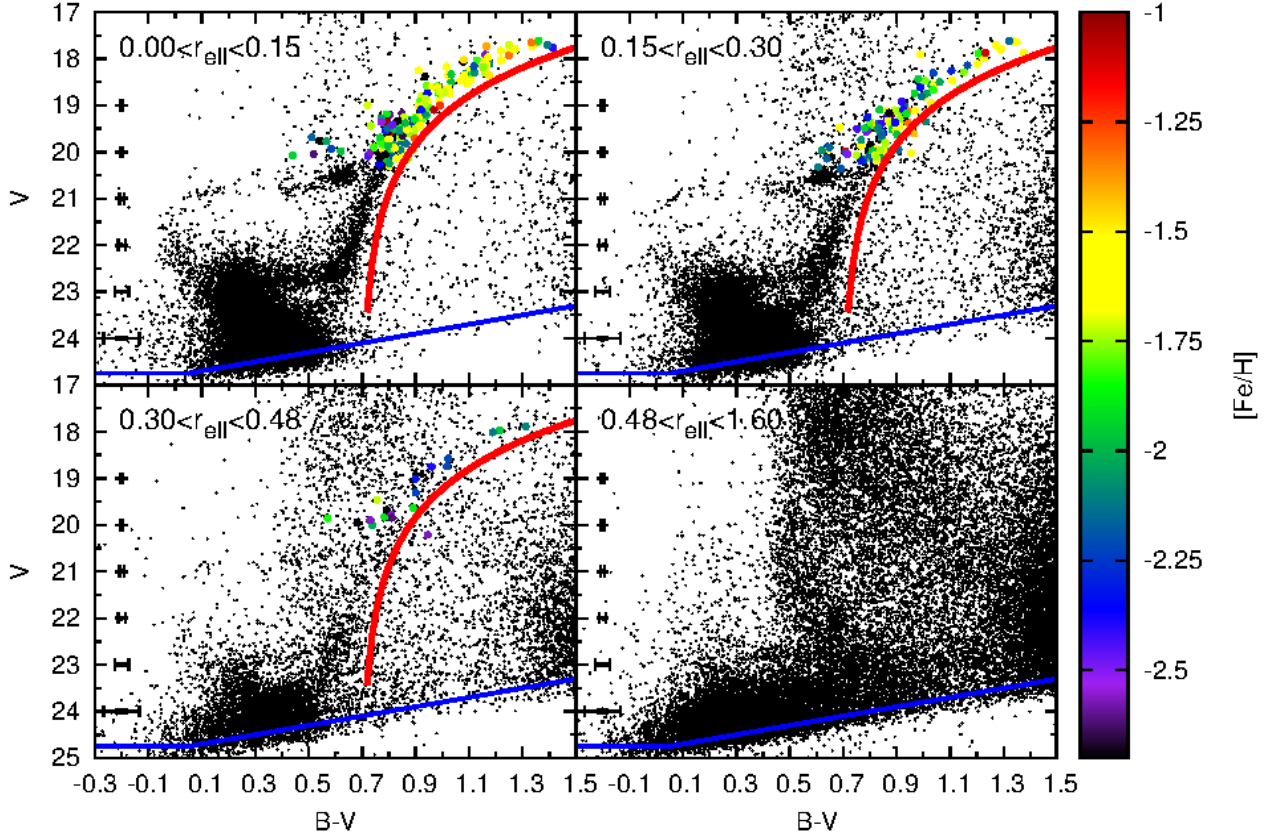}
\caption{The (V, B$-$V) CMDs of the Carina dSph at different radius from the centre, as shown in Figure~\ref{Carcov}. The [Fe/H] determinations of individual RGB stars from~\citet{Starkenburg10} are also shown as large coloured circles, with colour indicating the [Fe/H] abundance. The average photometric error at each magnitude level are represented by the horizontal error bars and the 50\% completeness level is indicated by the solid~(blue) line. Furthermore, the red line indicates the colour limit for lower RGB stars associated to Carina. \label{CarBVCaT}} 
\end{figure*}
\subsection{Photometry}
\label{photometry}
Deep optical photometry of the Carina dSph in the B and V filters was obtained using the CTIO 4-m MOSAIC II camera over 6 nights in December 2007 as part of observing proposal 2007B-0232~(PI M. Mateo), described in~\citet{Vivas13}. Several long~(600s), non-dithered exposures were obtained for each pointing and stacked together to obtain the deepest photometry possible. Short~(30s) exposures were also obtained for each pointing, to sample the bright stars that are saturated in the deep images. In order to ensure accurate photometric calibration, observations were made of Landolt standard fields covering a range of different airmass and colour~\citep{Landolt92, Landolt07}. \\
The spatial coverage of the deep photometric catalog is shown in Figure~\ref{Carcov}. The data are more than 95\% spatially complete within the tidal radius of Carina~(green, dashed ellipse in Figure~\ref{Carcov}) and extends well beyond the tidal radius out to a distance of 1.6 degrees along the major axis. \\
The photometric data were reduced and calibrated following the procedure outlined in~\citet{deBoer2011A}. Photometry was carried out using DoPHOT~\citep{Schechter93}. Observations of standard star fields were used to determine an accurate photometric calibration solution, depending on airmass, colour and brightness. Finally, the different fields were combined to obtain a single, carefully calibrated photometric catalog, which is given online in Table~1. Figure~\ref{Carphoterr} displays the photometric error as a function of magnitude for each filter, for all stars within the tidal radius.  \\
Figure~\ref{CarBVCaT} shows (V, B$-$V) CMDs of the Carina dSph at different radii from the centre. The average photometric error at each magnitude is indicated by the error bars and the 50\% completeness level is indicated by the solid~(blue) line. We resolve the oldest MSTO in Carina at V$\approx$23.5 with a photometric error of 0.025 mag in V and 0.05 in B$-$V, with a completeness of $\ge$80\%.

\subsection{Spectroscopy}
\label{spectroscopy}
Low resolution~(R$\sim$6500) \ion{Ca}{ii} triplet spectroscopy is available for 320 individual RGB stars in the Carina dSph, from VLT/FLAMES observations~\citep{Koch06, Helmi06} with the \ion{Ca}{ii} triplet [Fe/H] calibration from~\citet{Starkenburg10}. These observations provide [Fe/H] determinations for stars out to 0.5 degrees from the centre of the Carina dSph, and include a range in metallicities from $-$4.0$<$[Fe/H]$<$$-$1.0 dex. \\
HR spectroscopy from VLT/FLAMES is also available for individual stars, out to 0.3 degrees~\citep{Shetrone03, Koch08,Lemasle12,Venn12}. From these spectra, the metallicity [Fe/H] of 35 individual RGB stars is determined, and for some also the abundances of $\alpha$-elements~(O, Mg, Ca, Si, Ti) and r- and s-process elements~(Y, La, Ba, Eu, Nd). The HR spectroscopy covers a range in metallicity from $-$3.0$<$[Fe/H]$<$$-$1.2 dex.  \\
Figure~\ref{Carcov} shows the spatial coverage of the low resolution \ion{Ca}{ii} triplet~(open blue circles) and HR~(solid red dots) spectroscopy in comparison to the deep photometric data.

\section{SFH Method}
\label{SFHmethod}
The SFH of the Carina dSph will be determined using the routine Talos, which employs the synthetic CMD method~\citep[e.g.,][]{Tosi91,Tolstoy96,Gallart962,Dolphin97,Aparicio97}. This technique determines the SFH by comparing the observed CMDs to a grid of synthetic CMDs using Hess diagrams~(plots of the density of observed stars), taking into account photometric error and completeness. Uniquely, Talos simultaneously takes into account the photometric CMD as well as the spectroscopic MDF, providing direct constraints on the metallicity of stellar populations to obtain a well constrained SFH. We refer the reader to~\citet{deBoer2012A} for a detailed description of the routine as well as tests of its performance. 

\subsection{General setup}
\begin{figure}[!ht]
\centering
\includegraphics[angle=0, width=0.49\textwidth]{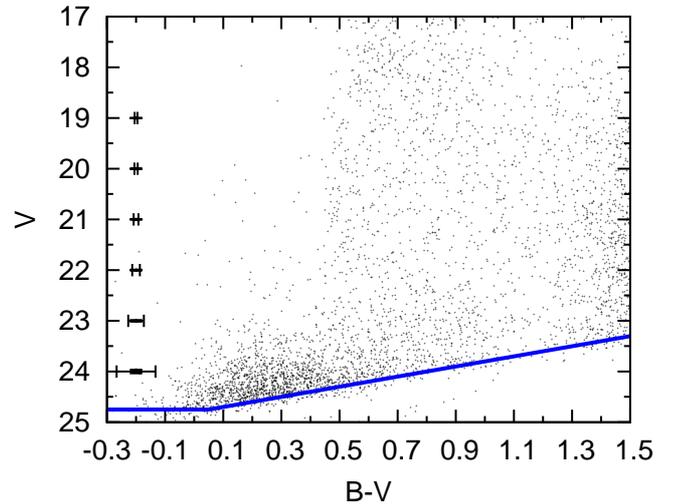}
\caption{Observed V,B$-$V CMD of the region adopted for foreground MW correction~(see Figure~\ref{Carcov}). Average photometric error is represented by the horizontal error bars and the 50\% completeness level is indicated by the solid blue line. \label{CarFG}}
\end{figure}

The CMDs of the outer regions of Carina contain a significant amount of contamination from stars belonging to the MW, as well as contamination from unresolved background galaxies which are recovered as single stars. During the SFH fitting, these stars will be incorrectly fitted using models at the distance of Carina, leading to anomalously old and/or metal-rich populations in the SFH solution. To correct for the presence of these objects, we adjust the photometric CMD by subtracting a representative ''empty" CMD. The correction region is chosen well outside the tidal radius of Carina~(see Figure~\ref{Carcov}) and should therefore be dominated by stars belonging to the MW and unresolved background galaxies. Figure~\ref{CarFG} shows that the CMD of the region is empty of features corresponding to Carina populations, and dominated by foreground MW stars with B$-$V$\ge$0.5 and background galaxies which are found mostly below V=23.5. The CMD of the correction region is scaled to the same size as the region under study and subtracted from the photometric CMD to correct for the presence of MW foreground stars and unresolved background galaxies. \\
To further reduce the effects of foreground contamination on the SFH, we also place a limit on the reddest colour of stars associated to Carina. Figure~\ref{CarBVCaT} shows the CMDs of Carina, overlaid with a line indicating the colour limit used as a function of brightness. The colour limit is chosen to be offset red ward from the lower Carina RGB at V=21 by four times the largest bin size adopted during the SFH determination~(0.1 in B$-$V). Stars on the lower RGB and MSTO with colours red ward of the limit are assumed to belong to the MW foreground, and disregarded during the SFH fitting. 
\\
Furthermore, previous photometric studies of Carina have revealed the presence of a gradient in the foreground extinction across the extent of Carina~\citep{Vivas13}. This varying foreground extinction can lead to artificial broadening of stellar evolution features, especially when simultaneously analysing data covering a wide spatial extent, such as our outer regions. Therefore, we interpolate within the dust extinction maps of~\citet{Schlegel98} to determine the extinction toward each individual star within our catalogues and create extinction-free CMDs\footnote{We used the tools available at the NASA/ IPAC Infrared Science Archive (http://irsa.ipac.caltech.edu/applications/DUST/)}. 

\begin{figure}[!ht]
\centering
\includegraphics[angle=0, width=0.49\textwidth]{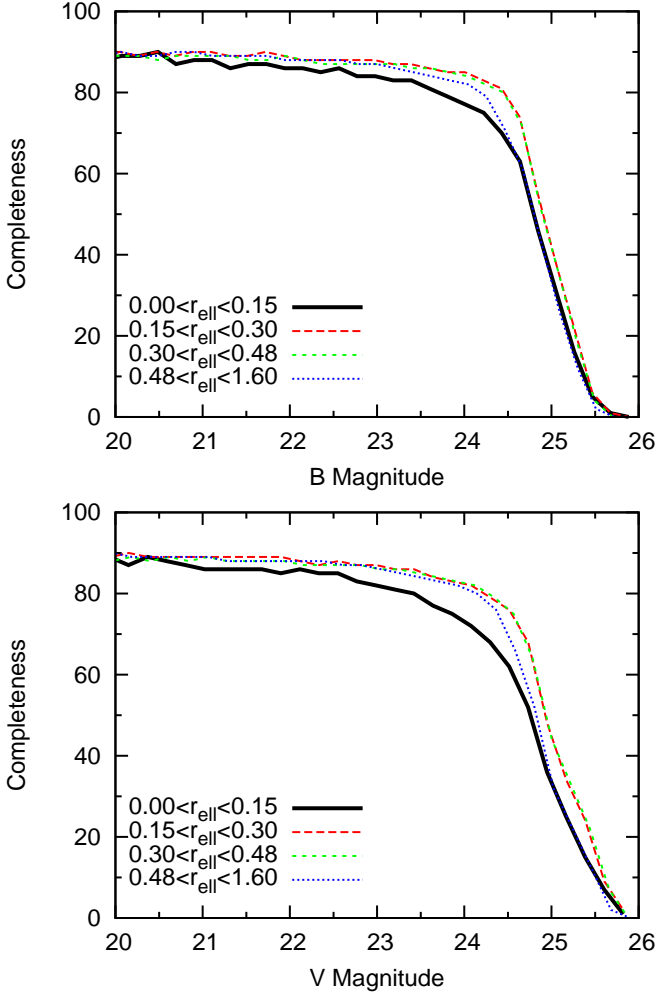}
\caption{Photometric completeness for the B and V filters for each annulus considered during the SFH determination, as determined from artificial star test experiments. \label{Carcomp}}
\end{figure}
\begin{figure}[!ht]
\centering
\includegraphics[angle=0, width=0.49\textwidth]{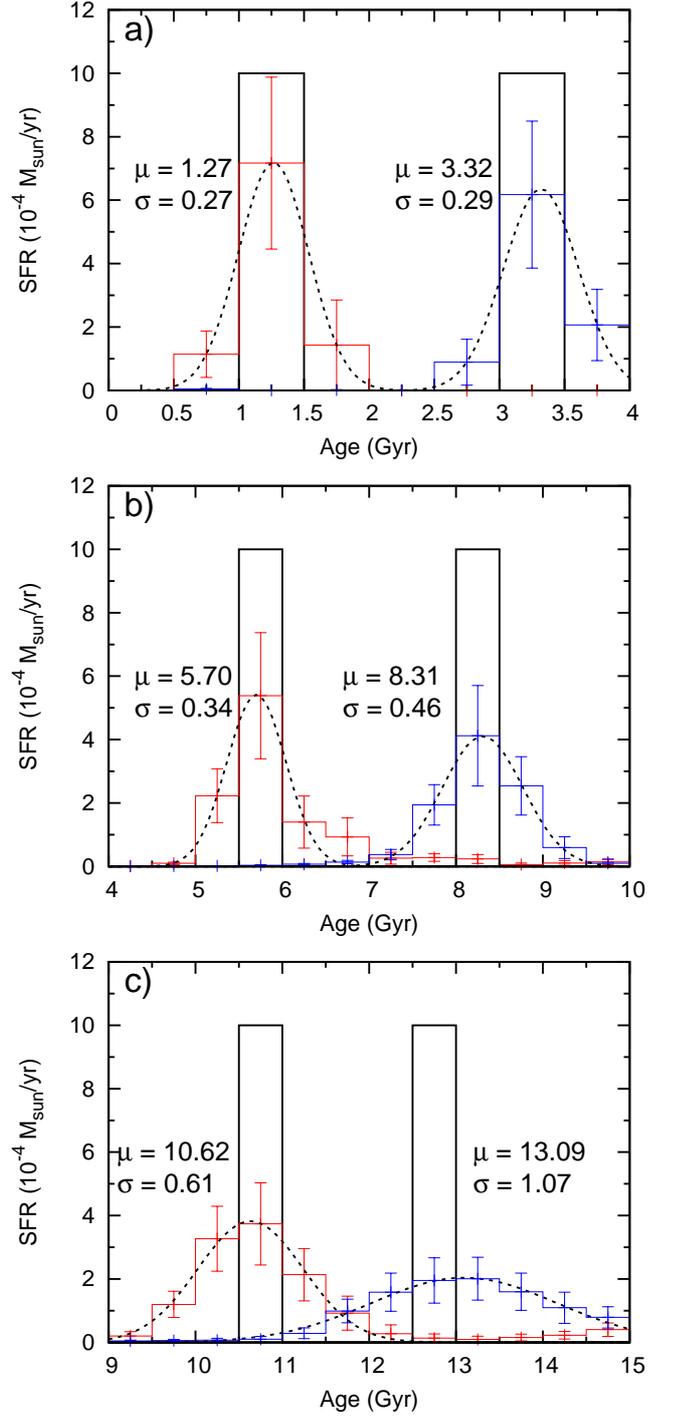}
\caption{The SFH of a series of synthetic short~(10 Myr) episodes of star formation at young~(a), intermediate~(b) and old~(c) ages. For each burst, the black histogram shows the input SFH, given the adopted binning of the solution. The coloured histograms show the recovered SFH of the burst, as well as the fit of a Gaussian distribution as the dashed line. The mean~($\mu$) and variance~($\sigma$) of the fitted Gaussian distribution are also listed. \label{CarSFHresolution}} 
\end{figure}

\subsection{Artificial star tests}
The observational conditions in the synthetic CMDs are simulated by using the results of extensive artificial star test simulations. This approach is the only way to take into account the complex effects that go into the simulation of observational biases, such as colour-dependence of the completeness level and the asymmetry of the photometric errors of stars at faint magnitudes~\citep[e.g.][]{Gallart961}. Artificial stars with known position and brightness were distributed randomly across the deep MOSAIC images of Carina, in both the B and V filters. No more than 5\% of the total observed stars were ever injected as artificial stars at one time, to avoid changing the crowding properties in the images. To obtain a robust determination of the completeness and photometric error in each part of the CMD, a total of 400 images with 5000 artificial stars was generated for each of the two inner fields and a total of 500 images with 2000 artificial stars each for the 4 outer fields. This resulted in 2800 images, containing a total of 8 million artificial stars spread across the full area of Carina.  \\
Figure~\ref{Carcomp} shows the photometric completeness of the data in the B and V filters at different distance from the centre of Carina. The completeness in each region is comparable due to the similar exposure times between fields, with slightly worse completeness fractions only in the centre of Carina due to stellar crowding. In particular, the completeness outside the tidal radius is very similar to the completeness inside the Carina tidal radius, showing that the effects of spatially varying completeness are negligible. Furthermore, Figure~\ref{Carcomp} also shows that the completeness of the foreground region used to correct for MW foreground contamination is comparable to that of regions inside the tidal radius. Therefore, subtracting the foreground region will not leading to systematic changes in the completeness in there resulting Hess diagram. 

\subsection{Age resolution}
\label{Carageresolution}
To obtain the age resolution of the SFH solution that can be achieved using the deep V,B$-$V photometry, we determine the ability of Talos to recover the age of a series of synthetic populations at different input ages~\citep[e.g.,][]{Hidalgo11}. We generated a set of six synthetic bursts of star formation at different ages covering the total age range we will study. Each burst has a duration of 10 Myr, and can therefore optimally be fit using one population bin in the SFH. However, due to the method of determining the uncertainties of the solution~(see Section~\ref{SFHmethod}), the solution for each burst will be smoothed across multiple bins of the SFH, affecting the resolution of the SFH at different ages. \\
The SFH for the synthetic episodes is recovered by fitting the V,B$-$V photometry simultaneously with a synthetic 50\% complete MDF with similar photometric depth as the~\ion{Ca}{ii} triplet spectroscopy. To determine the resolution of each episode, we fit a Gaussian distribution and determine the age of the central peak~($\mu$) as well as the variance~$\sigma$. 
\begin{figure}[!ht]
\centering
\includegraphics[angle=0, width=0.49\textwidth]{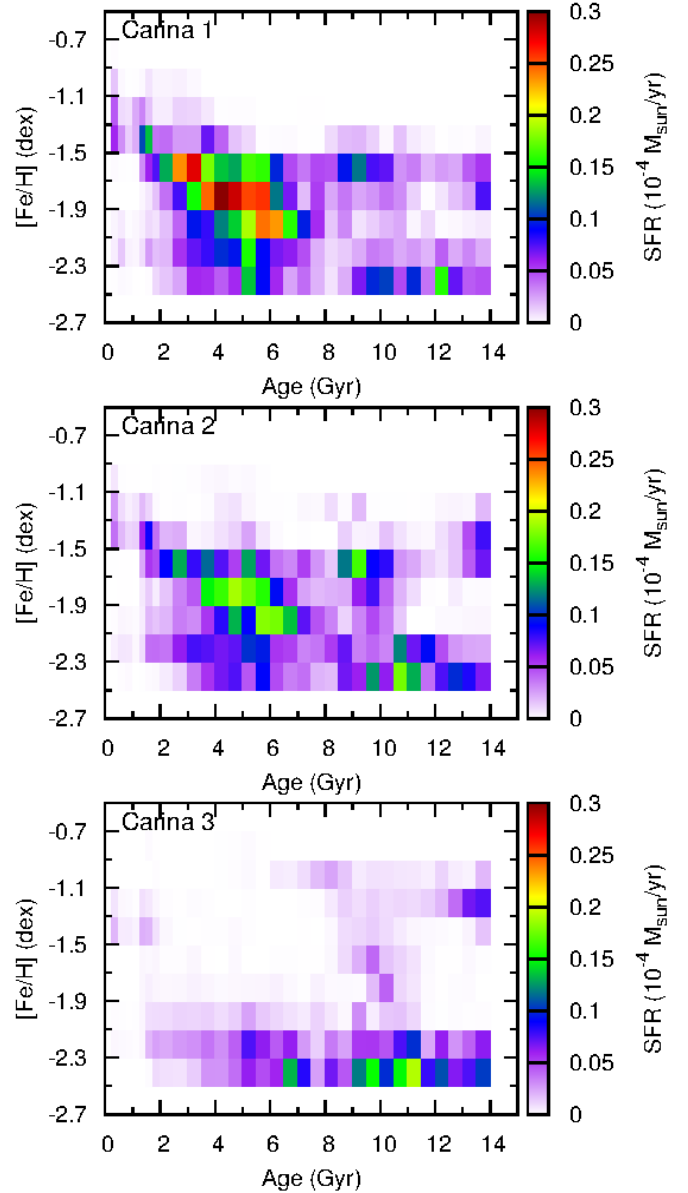}
\caption{The full SFH solution as a function of age and metallicity for the 3 different annuli~(r$<$r$_{core}$;  r$_{core}$$<$r$<$2 r$_{core}$; 2 r$_{core}$$<$r$<$r$_{tidal}$) within the Carina dSph, as well as the total SFH within the tidal radius. \label{Car123SFR}} 
\end{figure}
\\
The recovered and input SFH for the synthetic episodes is shown in Figure~\ref{CarSFHresolution}, for young~(a), intermediate~(b) and old~(c) star formation episodes. \\
For all but the oldest burst, the central peak is recovered within the correct bin. For the older age the peak of the synthetic episode is recovered at slightly too old ages. Figure~\ref{CarSFHresolution} shows that the age resolution of the recovered episode is significantly better at young ages than at old ages. For the young ages with small age resolution, the star formation is confined mostly to the central bin, while for the old episodes the star formation is spread out over multiple bins. The episodes are recovered with a resolution of~$\approx$0.25 Gyr at ages of 1.25 Gyr, 3.25 Gyr, 5.75 Gyr, $\approx$0.5 Gyr at ages 8.25 Gyr, 10.75 Gyr and~$\approx$1 Gyr at an age of 12.75 Gyr, which is consistent with values between 5 and 25\% of the adopted age.

\section{The Star Formation History of the Carina dSph}
\label{CarSFH}
To determine the SFH of Carina, we assume a large range of possible ages and metallicities, to avoid biasing the solution by our choice of parameter space. For the metallicity, a lower limit of [Fe/H]=$-$2.5 dex is assumed, which is the lowest available in our set of isochrones. The spectroscopic MDF shows that only $\approx$7 percent of stars have [Fe/H]$\le$$-$2.5 dex and therefore our lower metallicity limit should not lead to a bias in the SFH results. Furthermore, the metallicities of stars in the MDF also show that no stars with [Fe/H]$\ge$$-$1.0 dex are present on the RGB. However, we adopt an upper metallicity limit of [Fe/H]=$-$0.5 dex since higher metallicities may be present in young Main Sequence stars without a corresponding RGB sequence. A binsize of 0.2 dex is assumed for [Fe/H], which is similar to the average observed uncertainty on [Fe/H]. For the age limits, we assume a maximum age of 14 Gyr, for the age of the Universe, and consider a range of ages between 0.25~(the lowest age available in the isochrone sets used) and 14 Gyr old, with a binsize of 0.5 Gyr above an age of 2 Gyr and a bin size of 0.25 Gyr below 2 Gyr. 
\begin{figure*}[!ht]
\centering
\includegraphics[angle=0, width=0.99\textwidth]{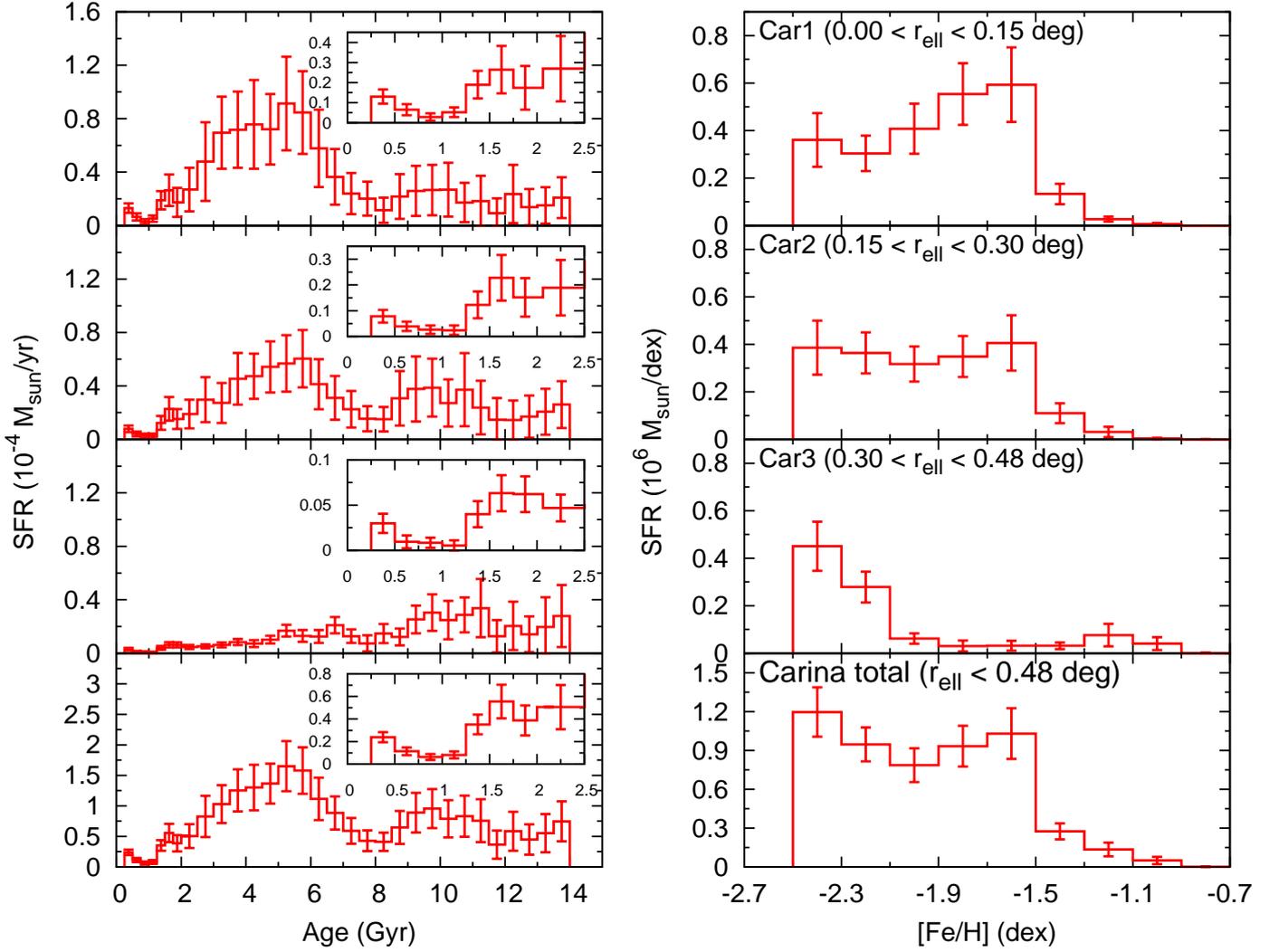}
\caption{The SFH~(left) and CEH~(right) of the 3 different annuli within the Carina dSph, with increasing distance from the centre. The radial extent of the annulus is indicated in each panel. The bottom row shows the total SFH and CEH of the Carina dSph within the tidal radius~(0.48 degrees). \label{Car123SFH}} 
\end{figure*}
\begin{figure*}[!ht]
\centering
\includegraphics[angle=0, width=0.99\textwidth]{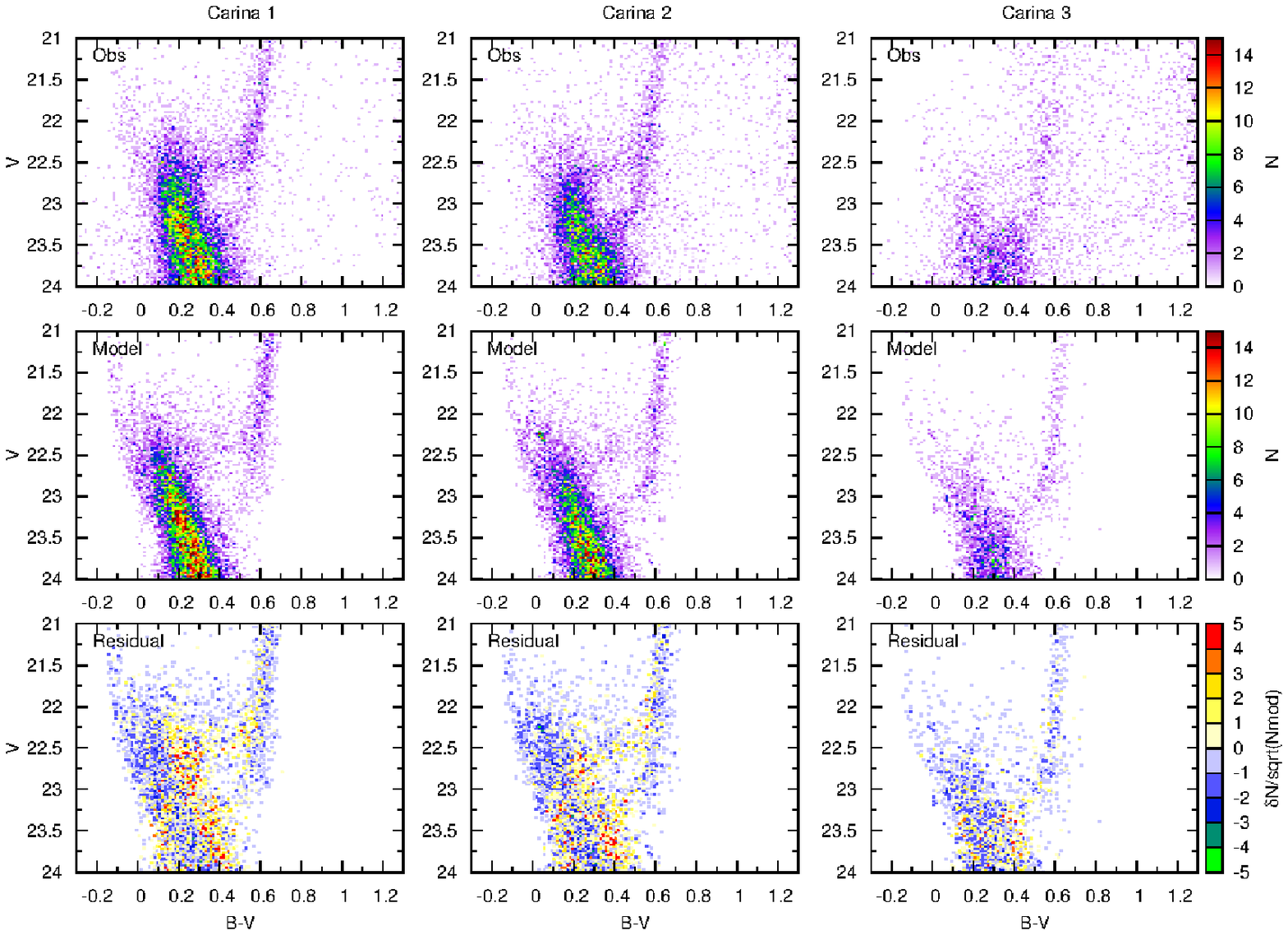}
\caption{The observed~(top row) and best-fitting~(middle row) CMD for the different annuli within Carina. Furthermore, the bottom row shows the difference between the observed and best-fit CMD, expressed as a function of the uncertainty in each CMD bin. \label{Car123residual}} 
\end{figure*}

\subsection{Star Formation History inside the tidal radius}
\begin{figure}[!ht]
\centering
\includegraphics[angle=0, width=0.49\textwidth]{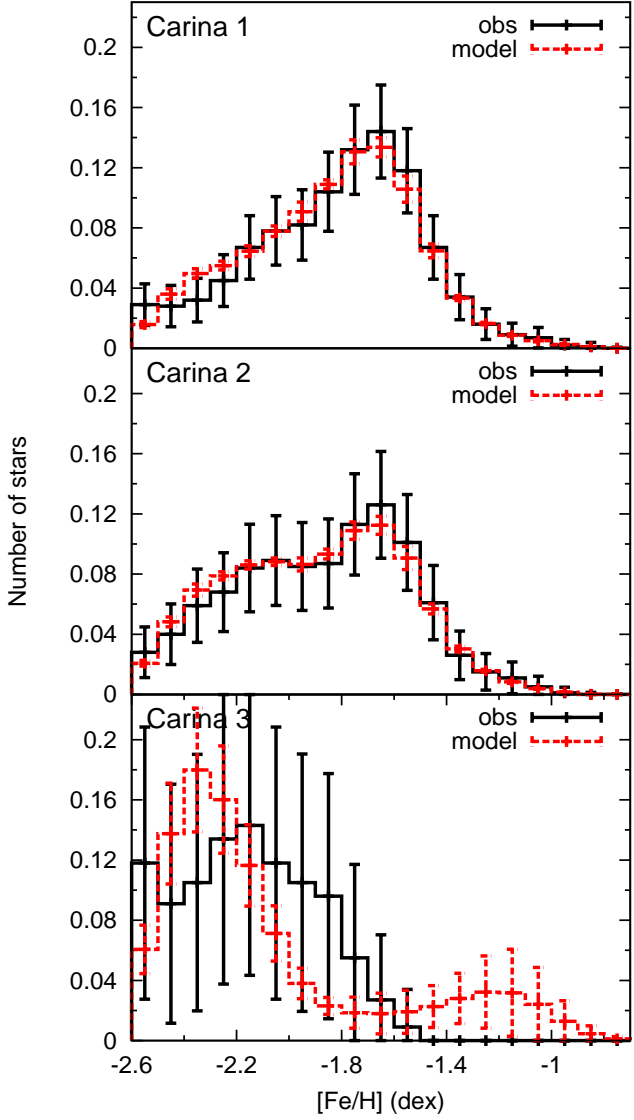}
\caption{Normalised metallicity distribution function for each of the three annuli within the tidal radius of Carina. The observed MDF is shown as black lines, while the best fit result from the SFH derivation is shown in red. The error bars indicate the uncertainties on the MDF due to uncertainties on the SFH for the model and due to Poissonian errors for the observations. \label{Car123MDF}} 
\end{figure}

Figure~\ref{Car123SFR} presents the final SFH solutions of the Carina dSph within the tidal radius for increasing distance from the centre~(r$<$r$_{core}$;  r$_{core}$$<$r$<$2 r$_{core}$; 2 r$_{core}$$<$r$<$r$_{tidal}$), determined using isochrones from the Dartmouth library~\citep{DartmouthI}. SFH solutions derived with a different set of isochrones~(BaSTI/Teramo) are presented in Appendix~\ref{CarSFHteramo}. By projecting the SFR values onto one axis we obtain the SFR as a function of age~(SFH) or metallicity~(Chemical Evolution History, CEH), given in Figure~\ref{Car123SFH}. The star formation rates as a function of age and metallicity derived from the SFH solution are also given online in Tables~2 and~3. The error bars indicate the uncertainty on the SFR as a result of different CMD and parameter griddings~(as described in Section~\ref{SFHmethod}). The SFH and CEH display the rate of star formation at different ages and metallicities over the range of each bin, in units of solar mass per year or dex respectively. The total mass in stars formed in each bin can be determined by multiplying the star formation rates by the range in age or metallicity of the bin. \\
Figures~\ref{Car123SFR} and~\ref{Car123SFH} show that the SFH of Carina can be divided into two main episodes, with old~($>$8 Gyr) and intermediate~($<$8 Gyr) ages. Figure~\ref{Car123SFR} also shows the presence of a very young episode of star formation in Carina, with a drop in SFR at an age $\approx$1 Gyr followed by a subsequent increase of SFR at the youngest ages probed in our SFH between 0.25$-$0.5 Gyr. Comparison to Appendix~\ref{CarSFHteramo} show that the bimodality of the SFH is not a result of the isochrone set used, but instead driven by the two distinct MSTOs visible in Figure~\ref{CarBVCaT}. Furthermore, the SFHs in Figure~\ref{Car123SFH} are roughly consistent with earlier studies of Carina~\citep[e.g.,][]{Monelli03, Bono10}, but resolve the different episodes in much greater detail. \\
Within the old star formation episode, the oldest stars have low metallicity, followed by a gradual metal enrichment over time to [Fe/H]$\approx$$-$1.5 dex at $\approx$8.5 Gyr. Following the old star formation episode, there is a clear paucity in star formation $\approx$8 Gyr ago that lasts for roughly 1 Gyr~(see Figure~\ref{Car123SFH}). Subsequently, star formation continues at an age of 7.5 Gyr, forming stars continuously until $\approx$250 Myr ago and gradually enriching the metallicity to [Fe/H]$\approx$$-$1.0 dex. Surprisingly, the oldest stars in the intermediate age episode display metallicities substantially lower than the stars formed at the end of the old episode. This indicates that the intermediate age star formation episode formed from gas that was initially not as metal enriched as the gas present at the end of the older episode, consistent with the gas infall scenario proposed by~\citet{Lemasle12}. \\
Comparison between the SFH at different distances from the Carina centre shows that a population gradient is present as a function of radius~\citep{Battaglia12}. Within the core radius, the SFH is dominated by the intermediate age~($<$8 Gyr) episode, which makes up 76 percent of the total star formation within this radius. Further out from the centre, star formation occurs in both episodes with roughly equal strength~(61 percent in intermediate age episode), while the outermost annulus is dominated by star formation in the oldest episode~(34 percent in intermediate age episode). This is consistent with the change in strength of the two distinct SGBs in Figure~\ref{CarBVCaT} as a function of radius. \\
Figure~\ref{Car123residual} shows the observed and synthetic (V, B$-$V) CMD of each region of Carina, as well as the residual in each bin in terms of Poissonian uncertainties. The residuals show that the synthetic CMDs are consistent with the observed CMDs within 3 sigma in each bin, indicating an overall good fit of the data. The synthetic CMDs correctly reproduce all evolutionary features in the lower CMD, including the location and spread of the distinct SGBs and the presence of younger populations above the bright SGB. \\
The observed MDF for each spatial region is shown in Figure~\ref{Car123MDF}, in comparison to the synthetic MDFs inferred from the SFH solution. For the inner two regions, the synthetic MDFs give a good fit of the observed spectroscopic MDFs, consistent within the small Poissonian error bars of the observations. This good fit is expected, given that the MDF for these regions is used as an input in the SFH fitting. The model MDF of the third region does not provide as good a fit to the observed data, although still consistent within the Poissonian errors. However, the uncertainties on the observed MDF are large, due to the low number of stars~(15) available within this annulus. The peak of the synthetic MDF is placed at low metallicity, broadly consistent with the observations. Metal-rich stars~([Fe/H]$>$$-$1.4 dex) are also present in the model MDF, consistent with anomalous metal-rich, old populations fit to residual MW foreground stars~(see also Figure~\ref{Car123SFR}). 
\\
Using the results of the SFH fitting, it is possible to determine the total stellar mass formed over the duration of star formation of Carina. By integrating the SFR in each population bin, we determine that the total mass in stars formed in Carina is 1.07$\pm$0.08$\times$10$^{6}$ M$_{\odot}$ within the tidal radius of 0.48 degrees or 888 pc~(0.43$\pm$0.05$\times$10$^{6}$ M$_{\odot}$ within the half-light radius of 250 pc). Assuming L$_{V}$=2.4$\pm$1.0$\times$10$^{5}$ L$_{\odot}$ for Carina, we determine a stellar mass-to-light ratio within the half-light radius of 1.8$\pm$0.8 compared to a dynamical mass-to-light ratio of $\approx$14~\citep{Irwin95, Walker092}. \\
By dividing the SFH into two components using a cut at 8 Gyr, we determine that 0.40$\pm$0.05$\times$10$^{6}$ M$_{\odot}$ formed within the oldest episode within the tidal radius, compared to 0.67$\pm$0.06$\times$10$^{6}$ M$_{\odot}$ within the intermediate age episode. Therefore, 60$\pm$9 percent of the total number of stars formed within the intermediate age episode, roughly consistent with studies of the stellar number count in each SGB~\citep{Smecker-Hane94,Rizzi03}.

\subsection{Star Formation History outside the tidal radius}
\begin{figure}[!ht]
\centering
\includegraphics[angle=0, width=0.49\textwidth]{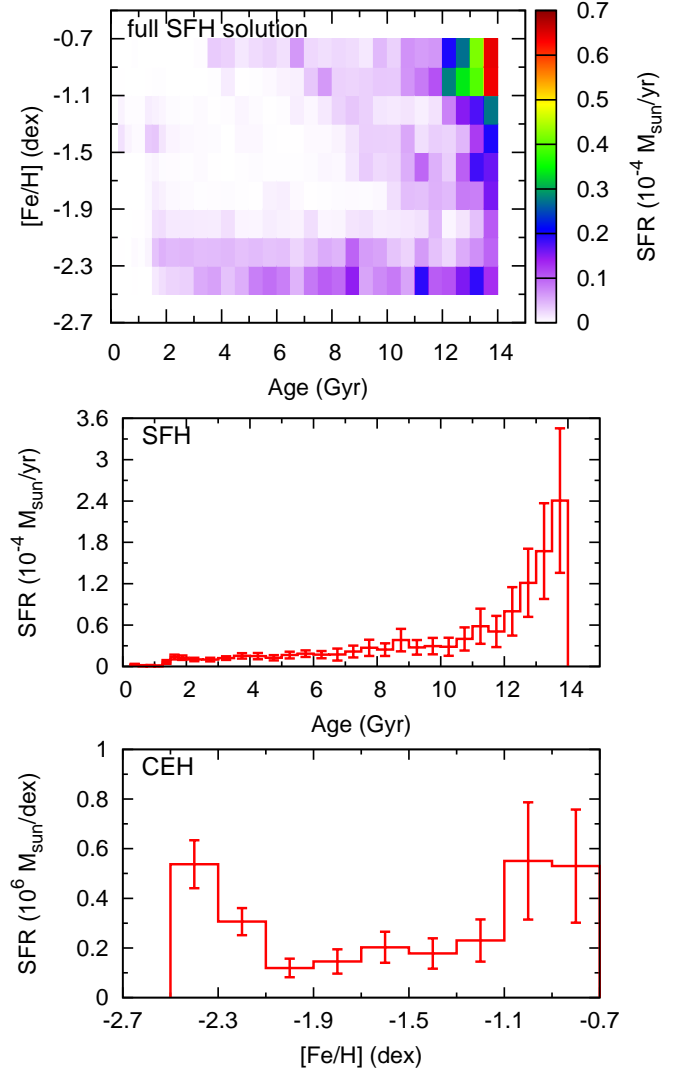}
\caption{The best-fitting SFH results for the stars in our sample outside the tidal radius of Carina. The top panel shows the full SFH solution as a function of age and metallicity, while the SFR as a function of age and metallicity are shown in the middle and bottom panels respectively.\label{Car4SFR}} 
\end{figure}

\begin{figure}[!ht]
\centering
\includegraphics[angle=0, width=0.49\textwidth]{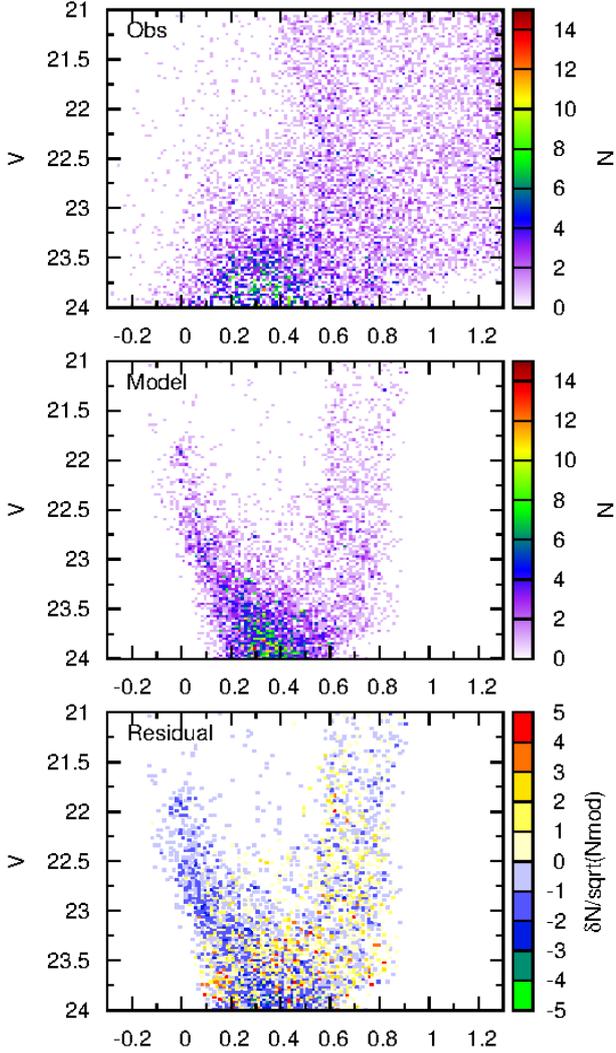}
\caption{The observed~(top) and best-fitting~(middle) CMD for the sample of Carina stars outside the tidal radius. The CMD residuals in each CMD bin are also shown in the bottom panel. \label{Car4residual}} 
\end{figure}

Figure~\ref{Car4SFR} presents the final SFH of Carina as a function of age and metallicity when considering only stars outside the tidal radius. Due to the significant MW foreground contamination in this region, no limit is placed on the reddest colour of stars associated to Carina. 
\\
The SFH presented in Figure~\ref{Car4SFR} is dominated by strong star formation at old ages~($>$10 Gyr) and high metallicities~([Fe/H]$>$$-$1.5 dex). This is consistent with fitting the red MW foreground stars using populations at the distance and extinction of Carina, and not representative of actual Carina populations. Excluding these populations, Figure~\ref{Car4SFR} shows the presence of metal-poor populations broadly consistent with those seen in Figure~\ref{Car123SFR} at $\approx$6 and $\approx$11 Gyr. Therefore, it seems plausible that the tidal tails detected around Carina are composed of populations similar to those found just within the tidal radius. However, the excessive MW foreground contamination prevents us from determining the properties of these populations in detail. \\
Figure~\ref{Car4residual} shows the observed and synthetic CMDs of the extra-tidal region, along with CMD residuals in terms of Poissonian uncertainties. A broad, double RGB is visible in the model CMD, which is dominated by populations fit to the MW foreground. The observed CMD shows hints of the old SGB at B$-$V=0.45, V=23.25, which are correctly reproduced in the synthetic CMD. This further hints at the presence of Carina stars outside the nominal tidal radius in the form of tidal tails.

\section{Chemical evolution timescale in the Carina dSph}
\label{Carindivages}
Using the SFH presented in Figures~\ref{Car123SFR} and~\ref{Car123SFH}, we determine the probability distribution function for age of individual stars on the RGB, taking into account the SFH constraints for the age at each CMD location. For each observed RGB star, all stars in the synthetic CMD satisfying the observed magnitude and metallicity~(and uncertainties) are used to build up the distribution of age at this CMD location, following the procedure outlined in~\citet{deBoer2012A}. The median of this age distribution is adopted as the age of the observed star, while the median absolute deviation~(MAD) is used as an error bar. Using this method, we determine accurate age estimates for all available samples of spectroscopic stars in Carina~\citep{Shetrone03, Helmi06,  Koch08, Starkenburg10,Lemasle12,Venn12}. 
\\
Using stars with spectroscopic metallicity determinations, we can determine the detailed Age-Metallicity Relation~(AMR) for individual RGB stars. Figure~\ref{Car123ages}a shows the AMR for low resolution~\ion{Ca}{ii} triplet spectroscopic samples~(red points) and high resolution spectroscopic samples~(blue points). Filled circles denote stars for which the probability distribution for age could be determined, while filled triangles show stars for which no statistical age estimate could be derived. For those stars, only an age estimate is given, based on the closest distance from synthetic CMD points satisfying the magnitude and metallicity constraints. Several stars in Figure~\ref{Car123ages}a display ages $\approx$1 Gyr, inconsistent with lifetimes associated to RGB stars. Inspection of these stars in the CMD shows that they are likely AGB stars instead of RGB stars, in which case classification as RGB stars would result in too young ages. \\
The AMR presented in Figure~\ref{Car123ages}a is broadly consistent with previous studies such as~\citet{Lemasle12}, but our results determine ages with greater accuracy due to the constraints from the full CMD analysis. Therefore, instead of merely separating stars into two episodes, we are now able to determine the detailed behaviour in each episode of star formation. Similar to~\citet{Lemasle12}, the AMR can be roughly divided into two parallel sequences using an age cut at 8 Gyr, although the gap in the SFH is not as visible as in Figure~\ref{Car123SFR}. The old stars form a narrow, well-defined AMR, very similar to that of the Sculptor dSph~\citep{deBoer2012A}. Stars with low metallicity are formed at ancient ages~($>$12 Gyr), followed by a phase of gradual metal enrichment during several gigayears up to [Fe/H]$\approx$$-$1.5 dex at an age of $\approx$8 Gyr. \\
Stars associated to the intermediate age episode mainly display metallicities between $-$2.0$\ge$[Fe/H$\ge$$-$1.5 dex with ages of 2$-$7 Gyr. A sequence of more metal-poor stars is also visible with ages of 4$-$6 Gyr. Finally, at young~($<$3 Gyr) ages, the metallicity of stars increases again, rising from [Fe/H]=$-$1.5 dex at $\approx$3 Gyr to [Fe/H]=-1.0 dex at $\approx$1 Gyr. Stars with metallicities even higher than this may also be present in Carina, based on the appearance of young, blue MS stars in Figure~\ref{CarBVCaT}. However, these populations cannot be properly sampled using spectroscopic observations of upper RGB stars, since they are present in large numbers only at fainter magnitudes. \\
Using stars from high resolution spectroscopic samples, we can determine the evolution of individual chemical elements as a function of time. Figures~\ref{Car123ages}b,c,d shows the distribution of [Mg/Fe], [Ca/Fe] and [Ba/Fe] as a function of metallicity, colour coded with the statistical age of the individual stars. Unfortunately, stars with low~([Fe/H]$<$$-$2.0 dex) metallicities in the intermediate age episode are not fully sampled in the HR spectroscopic sample, preventing us from seeing the behaviour of $\alpha$-element abundances during the onset of the second star formation episode. \\
When considering old~($>$8 Gyr) stars, Figure~\ref{Car123ages} shows a clear sequence of steadily declining $\alpha$-element ratios with metallicity, from metal-poor and MW-like to more metal-rich and [$\alpha$/Fe]$<$0, similar to what was seen in~\citet{Lemasle12}. This effect is more pronounced in [Mg/Fe] than in [Ca/Fe], but still present in Figure~\ref{Car123ages}c. If this decrease in [$\alpha$/Fe] is interpreted as a sign of the presence of an $\alpha$-element ``knee"~\citep{Tinsley79}, the location of the knee would be around [Fe/H]$\approx$$-$2.5 dex and ancient~($\approx$12 Gyr) age~\citep{Lemasle12}. Contrary to~\citet{Lemasle12}, the sequence formed by stars classified as old does not extend all the way down to [Fe/H]=$-$1.2 dex but stops at [Fe/H]=$-$1.5 dex and only slightly sub-solar $\alpha$-element abundances. At this point, the $\alpha$-element ratios increase for younger ages, leading to the distinct group of stars with [Fe/H]$\approx$$-$1.5 dex and [$\alpha$/Fe] consistent with MW abundances. These stars cover a range of ages between roughly 4-10 Gyr, and correspond to the metal-rich stars at the end of the old star formation episode and the middle of the intermediate age episode. Finally, the youngest stars in the HR dataset form a sequence of decreasing [$\alpha$/Fe] which extends from the group of intermediate age, MW-like stars down to $\alpha$-element abundances well below solar. \\
Additionally, one star from the group of intermediate age, metal-poor sequence is also present in the HR spectroscopic dataset, and can therefore be used to probe the abundances of stars associated to the onset of the second star formation episode. The $\alpha$-element abundances of this star are consistent with those of the MW at low metallicities, indicating that the event which triggered the onset of the second episode of star formation is likely also responsible for the decrease in metallicity and corresponding increase in [$\alpha$/Fe] ratios.

\section{Discussions and conclusions}
\label{conclusions}
\begin{figure*}[!ht]
\centering
\includegraphics[angle=0, width=0.45\textwidth]{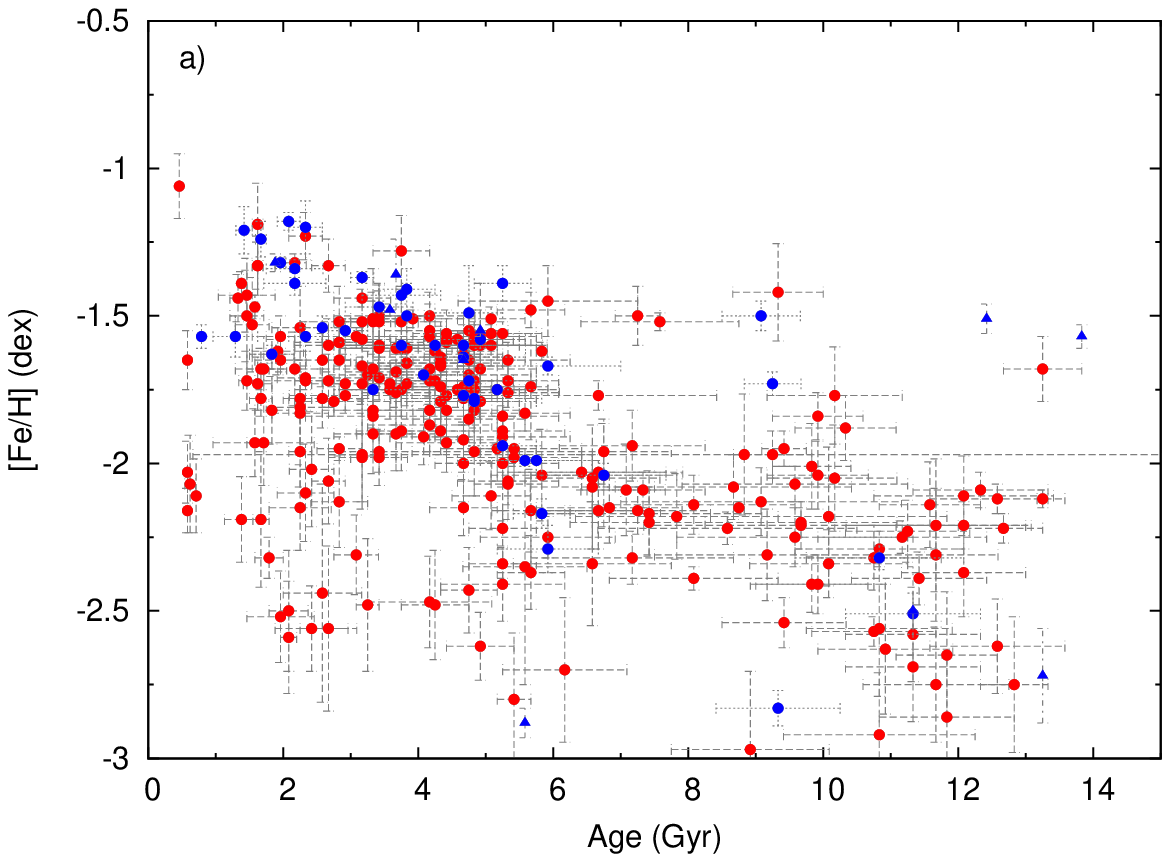}
\includegraphics[angle=0, width=0.45\textwidth]{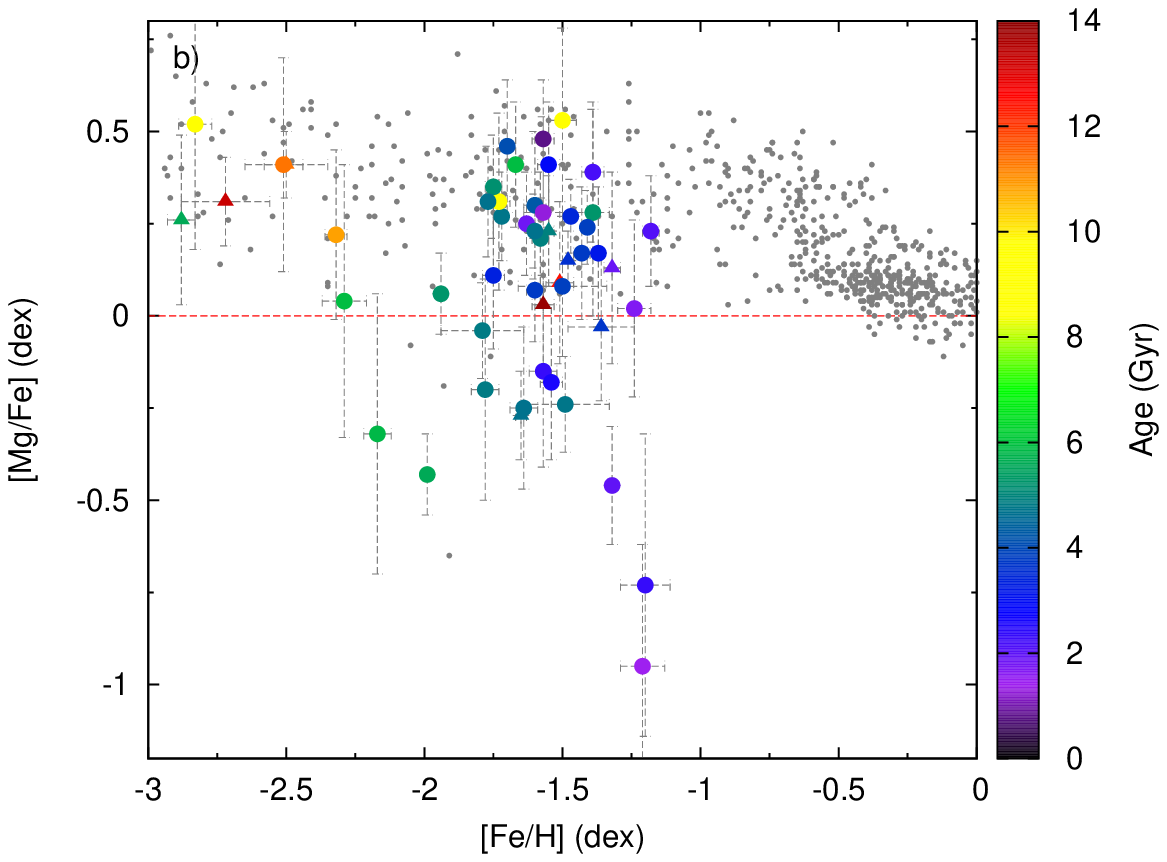}
\includegraphics[angle=0, width=0.45\textwidth]{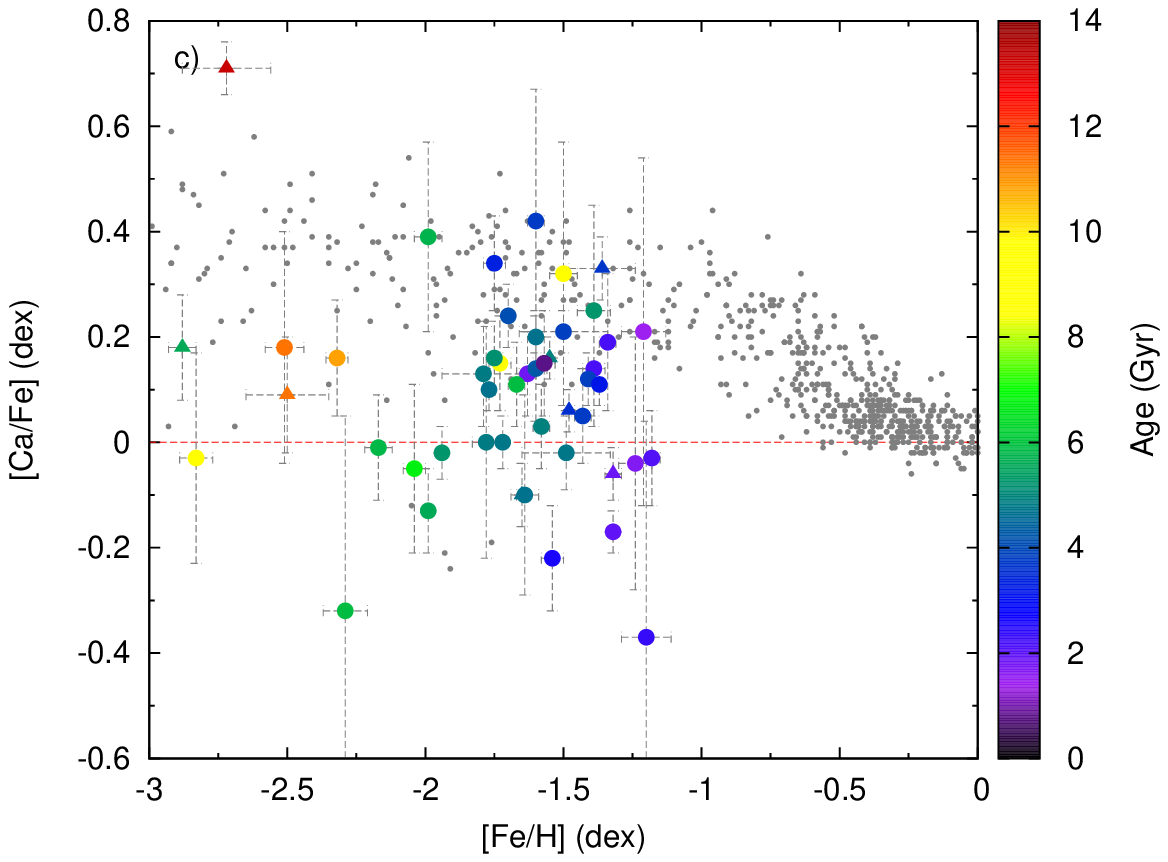}
\includegraphics[angle=0, width=0.45\textwidth]{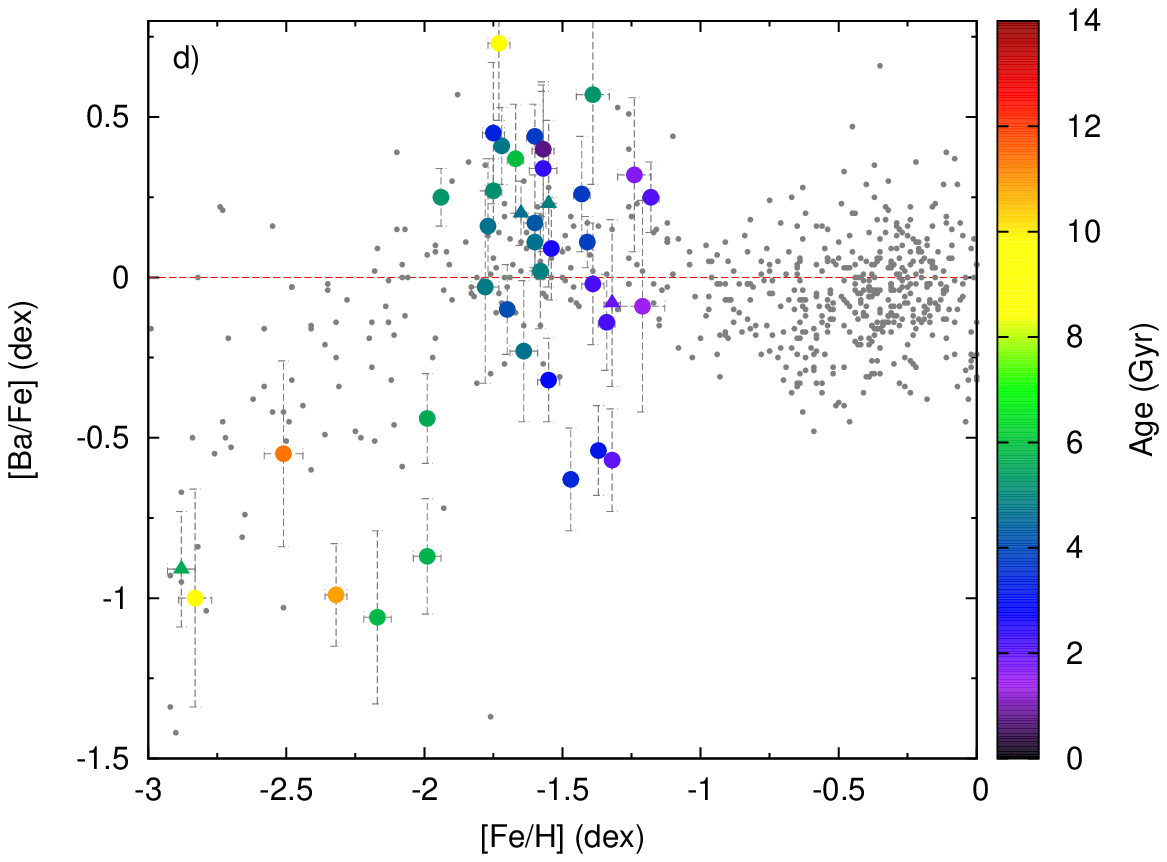}
\caption{The top-left panel shows the Age-Metallicity Relation of the Carina dSph as obtained from individual stars on the upper RGB. Low resolution~\ion{Ca}{ii} triplet spectroscopy from~\citet{Starkenburg10} are shown in red, while high resolution spectroscopy from~\citep{Shetrone03, Koch08,Lemasle12,Venn12} are shown as blue points. The [Mg/Fe]~\textbf{b)}, [Ca/Fe]~\textbf{c)} and [Ba/Fe]~\textbf{d)} abundances for the high resolution spectroscopic samples of RGB stars from~\citet{Shetrone03, Koch08,Lemasle12,Venn12} are also shown. Triangles indicate stars for which no statistical age estimate could be determined. For those stars, only a rough age is given, based on the closest distance from evolutionary features in the SFH. The colours represent the age estimate of individual stars in Gyr, as derived from the SFH. Stars in the Milky Way are shown for comparison~(small grey points). \label{Car123ages}} 
\end{figure*}

The Carina dSph has undergone a complex, episodic formation history leading to distinct SGBs in the photometric CMD~(see Figure~\ref{CarBVCaT}) and complex chemical abundance patterns. In this work, we have presented the detailed SFH and CEH of Carina at different positions within its tidal radius, using a combination of deep photometry and spectroscopic metallicities. The spatially resolved SFH~(see Figures~\ref{Car123SFR} and~\ref{Car123SFH}) confirms the episodic formation history of Carina and quantifies the strength and stellar population make-up of each episode. By combining the SFH results with observations of chemical abundances of individual RGB stars, we are also able to determine the changes in metallicity and chemical abundance patterns as a function of age~(see Figure~\ref{Car123ages}). \\
The mean ages of the dominant star formation episodes are consistent with previous studies~\citep[e.g.,][]{Monelli03, Bono10}, but the SFH presented here is able to determine, for the first time, the age and metallicity extent of each episode separately. The SFH of Carina is determined at different radii from the centre, confirming and quantifying the negative age and metallicity gradient as a function of radius~\citep{Harbeck01, Battaglia12}. Furthermore, the SFH also hints at the presence of Carina stellar populations outside of the tidal radius~(see Figure~\ref{Car4SFR}), which is consistent with the presence of tidal tails~\citep{Battaglia12}.  \\
The SFH presented in Figure~\ref{Car123SFR} points to a complex formation for Carina, characterised by two main episodes which both formed stars starting from low metallicities~([Fe/H]$<$$-$2 dex). The overlap in metallicity was first pointed out by~\citet{Lemasle12} from HR spectroscopic observations and now independently confirmed by fitting the deep MSTO photometry. This points to a different metal enrichment in each star formation episode, and possibly a different origin of the gas that formed the stars in each episode. The overlapping metallicity range of both episodes also explains the thin RGB and narrow MDF of Carina found in previous studies~\citep{Armandroff91, DaCosta94,Smecker-Hane96, Smecker-Hane99}. \\
The old stars~($>$8 Gyr) in Carina display a narrow sequence in Figure~\ref{Car123SFR} and a narrow AMR from~\ion{Ca}{ii} triplet and HR spectroscopic observations. Similar to~\citet{Lemasle12}, the old episode shows steadily declining $\alpha$-element abundances as a function of metallicity and age, consistent with evolution in relative isolation~(Figure~\ref{Car123ages}b,c). A possible $\alpha$-element ``knee" is visible at very low metallicities~(Fe/H]$\approx$$-$2.5 dex). The metallicity of the knee indicates that it experienced a slow metal enrichment before the onset of SNe Ia explosions, consistent with the low mass of Carina compared to other LG dSphs. \\
The stars associated to the intermediate age episode display a more complex behaviour as a function of age and metallicity. The SFH shows a broad sequence of star formation with increasing metallicity as a function of age, but with a large scatter in metallicity. The oldest star formation in the intermediate age episode peaks at [Fe/H]$\approx$$-$2.0 dex, in Figure~\ref{Car123SFR} roughly 0.5 dex more metal-poor than the stars formed at the end of the old episode. This is also reproduced in the AMR~(Figure~\ref{Car123ages}a), where the majority of stars with ages between 4$-$8 Gyr are more metal-poor than the stars at the end of the old sequence. Figure~\ref{Car123ages}b,c shows that stars from the intermediate age episode also display different abundance patterns, with significantly higher [$\alpha$/Fe] ratios~\citep{Lemasle12}. The stars form a sequence starting from [Fe/H]=$-$1.8 dex and [Mg/Fe]$\approx$0.4 dex and declining to Mg-poor values~( [Mg/Fe]$\leq$$-$0.5 dex) with a slope much steeper than observed in the old episode. This indicates clearly that both episodes of star formation formed from gas with different abundance patterns. 
\\
The complex SFH of Carina cannot easily be explained by a scenario involving evolution purely in isolation, unless a scenario of in substantial homogenous mixing is invoked. Instead, the spatial, chemical and temporal distributions of Carina stars point to an external event as the origin for the episodic formation history. An external event would also be consistent with simulations, which invoke tidal interactions with the MW to self-consistently model the formation of the Carina dSph~\citep{Pasetto11}. Orbit determinations of Carina do suggest that it has experienced multiple interactions with the MW, with close perigalacticon passages at 0.7, 2.1, 3.6, 5.0, 6.5, 7.9, 9.4, and 10.8 Gyr, broadly consistent with the presence of star formation at these ages in the SFH~\citep{Piatek03}. However, it is unclear how the effects of tidal interactions can have a major effect on the SFH, while still preserving the population gradients observed in Carina. \\
Gas infall is an attractive scenario to explain the complex formation history of the Carina dSph~\citep{Lemasle12}. Accretion of gas can explain the change in abundances as well as the renewed star formation activity. However, to change the [$\alpha$/Fe] ratios and metallicity of the ISM this gas could not have been substantially pre-enriched. If the infall of fresh gas did indeed lead to a ``restart" of chemical enrichment, the distribution of $\alpha$-element should show a second sequence starting at low metallicities, and possibly a second $\alpha$-element ``knee". Unfortunately, the current sample of HR spectroscopic observations of stars consistent with the onset of the intermediate is limited to only one star. \\
It is still unclear what has led to the paucity in star formation between both episodes at an age of $\approx$8 Gyr. Figure~\ref{Car123SFH} does not show a clear drop in SFR as a function of age in the old episode, which makes it unlikely that gas depletion was the reason for the initial paucity in star formation. Furthermore, if the initial star formation in the old episode led to the blowout of gas and eventual cessation of star formation, it is hard to conceive how the intermediate episode was able to form stars at an overall higher rate~(60$\pm$9 percent of stars within the intermediate age episode) while holding on to its gas long enough to keep forming stars for $\approx$7 Gyr more. If Carina was able to retain more gas in the intermediate age episode than in the old episode, this could point to the addition of more~(dark) mass along with the gas that triggered the renewed star formation, such as through the accretion of a dark halo or dwarf galaxy. \\
The complex, episodic history of Carina  provides a unique window into dwarf galaxy evolution dominated by external processes, and provides a challenging testbed for any theory of galaxy formation.  

\section{Acknowledgements}
\label{acknowledgements}
The research leading to these results has received funding from the European Research Council under the European UnionÕs Seventh Framework Programme (FP/2007-2013) / ERC Grant Agreement n. 308024. T.d.B. acknowledges financial support from the ERC. G.B. is grateful to the International Space Science Institute~(ISSI), Bern, Switzerland, for supporting and funding the international team "First stars in dwarf galaxies".

\bibliographystyle{aa}
\bibliography{Bibliography}

\clearpage
\begin{appendix}
\onecolumn
\section{SFH solution using BaSTI isochrones}
\label{CarSFHteramo}
\begin{figure*}[!ht]
\centering
\includegraphics[angle=0, width=0.99\textwidth]{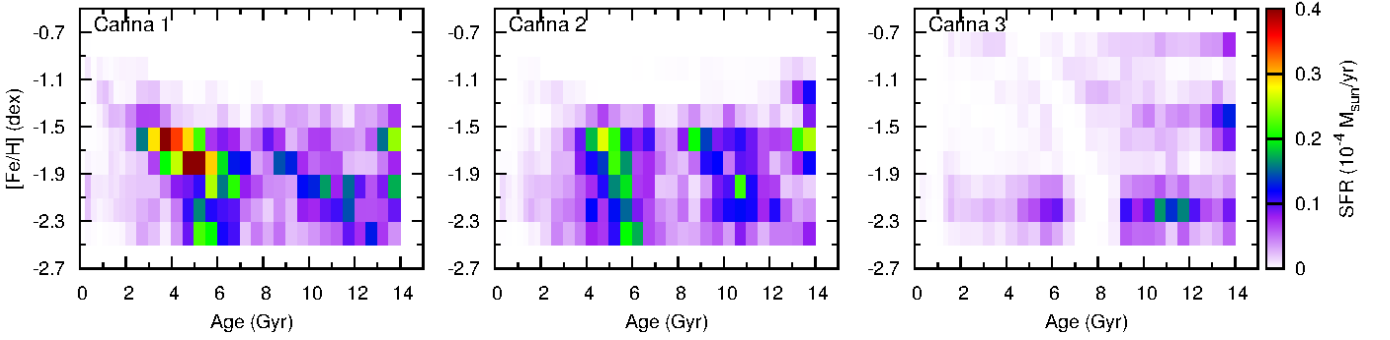}
\caption{The full SFH solution as a function of age and metallicity for each of the 3 annuli within the Carina dSph, when adopting the Teramo isochrone set. \label{Car123SFR_TER}} 
\end{figure*}

\begin{figure*}[!ht]
\centering
\includegraphics[angle=0, width=0.8\textwidth]{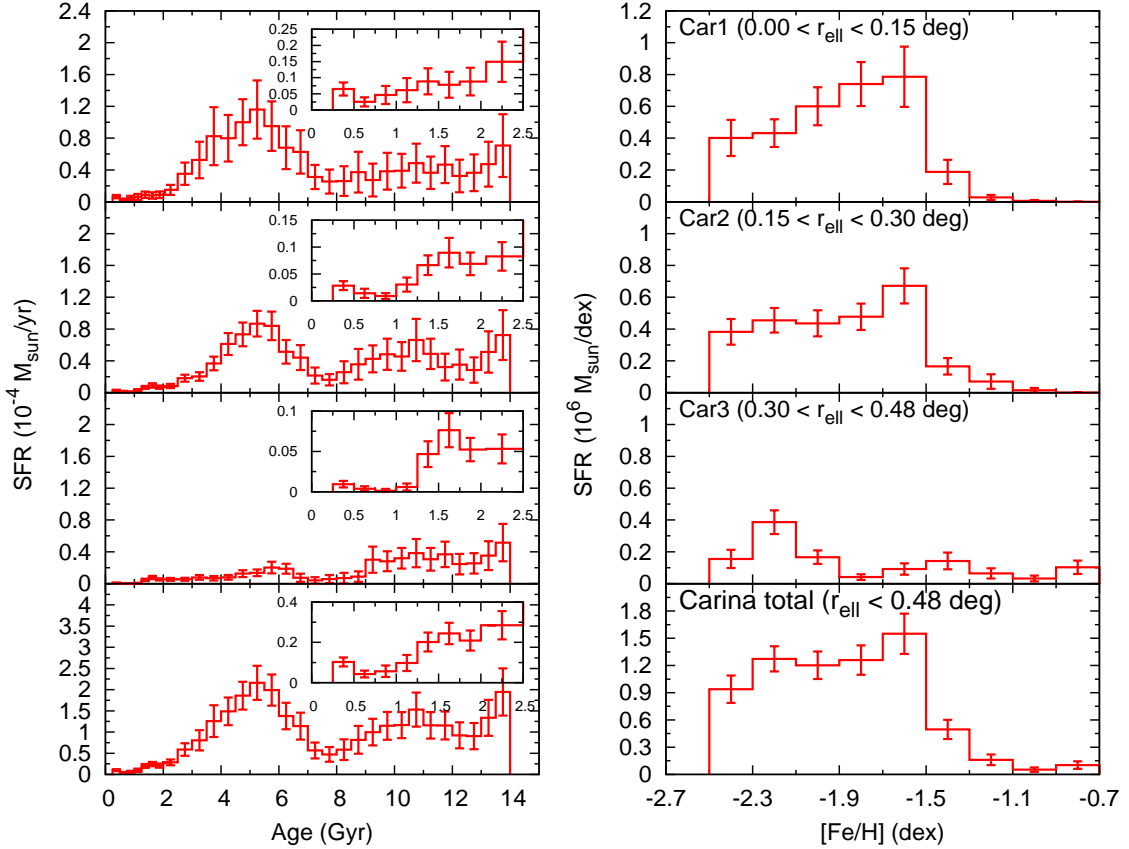}
\caption{The SFH~(left) and CEH~(right) of the 3 different annuli within the Carina dSph, when adopting the Teramo isochrone set.\label{Car123SFH_TER}} 
\end{figure*}

\begin{figure*}[!ht]
\centering
\includegraphics[angle=0, width=0.695\textwidth]{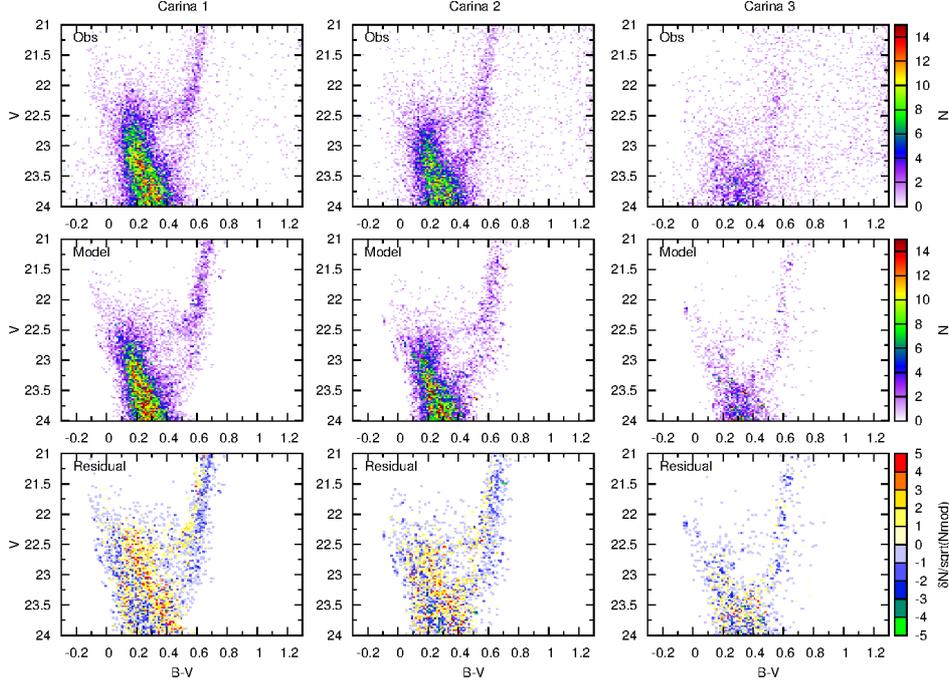}
\caption{The observed~(top row) and best-fitting~(middle row) CMD for the different annuli within Carina, when adopting the Teramo isochrone set. The bottom row shows the difference between the observed and best-fit CMD, expressed as a function of the uncertainty in each CMD bin. \label{Car123residual_TER}} 
\end{figure*}

\begin{figure*}[!ht]
\centering
\includegraphics[angle=0, width=0.35\textwidth]{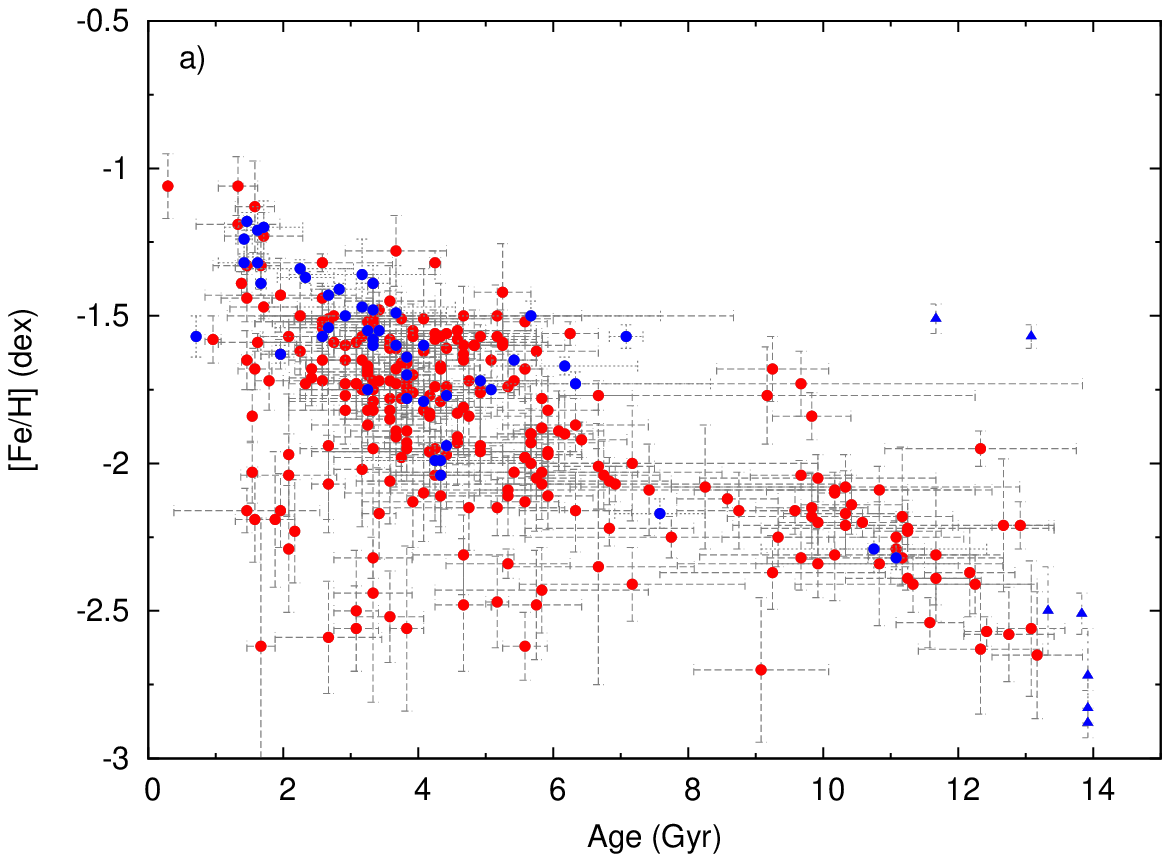}
\includegraphics[angle=0, width=0.35\textwidth]{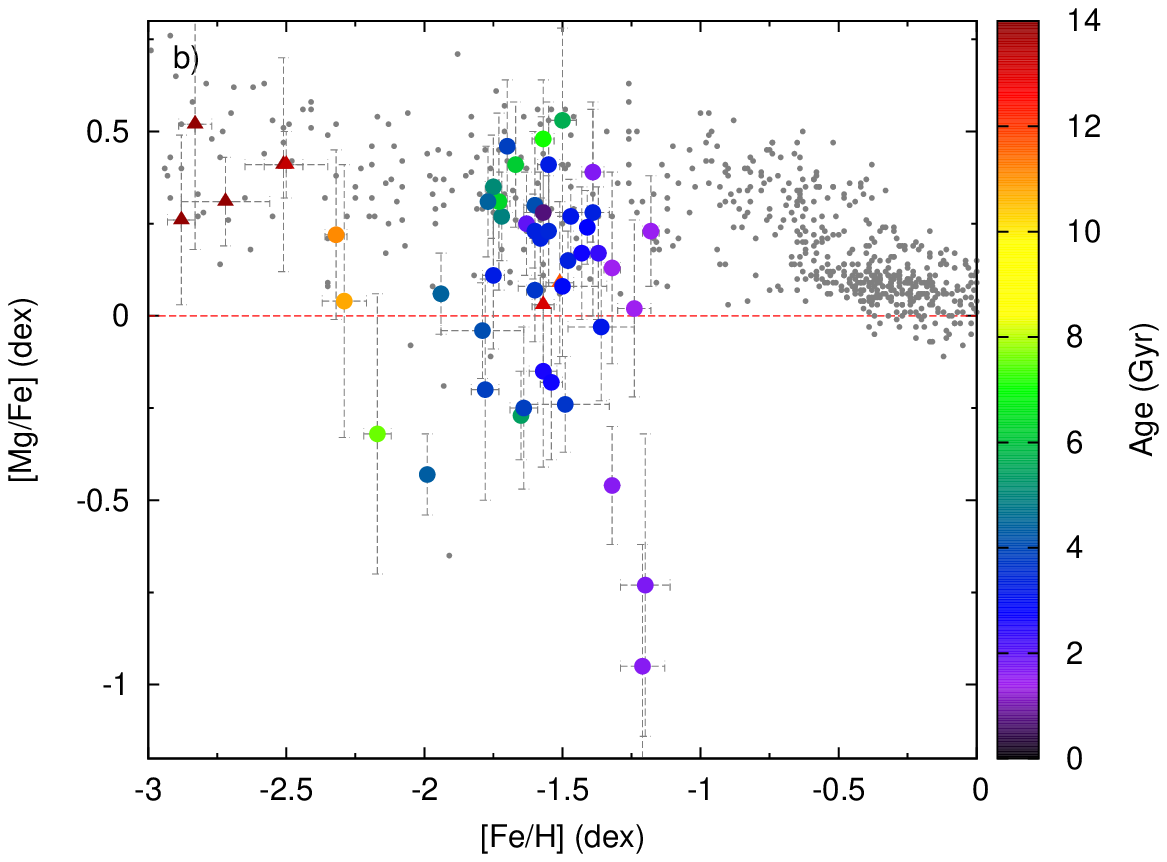}
\includegraphics[angle=0, width=0.35\textwidth]{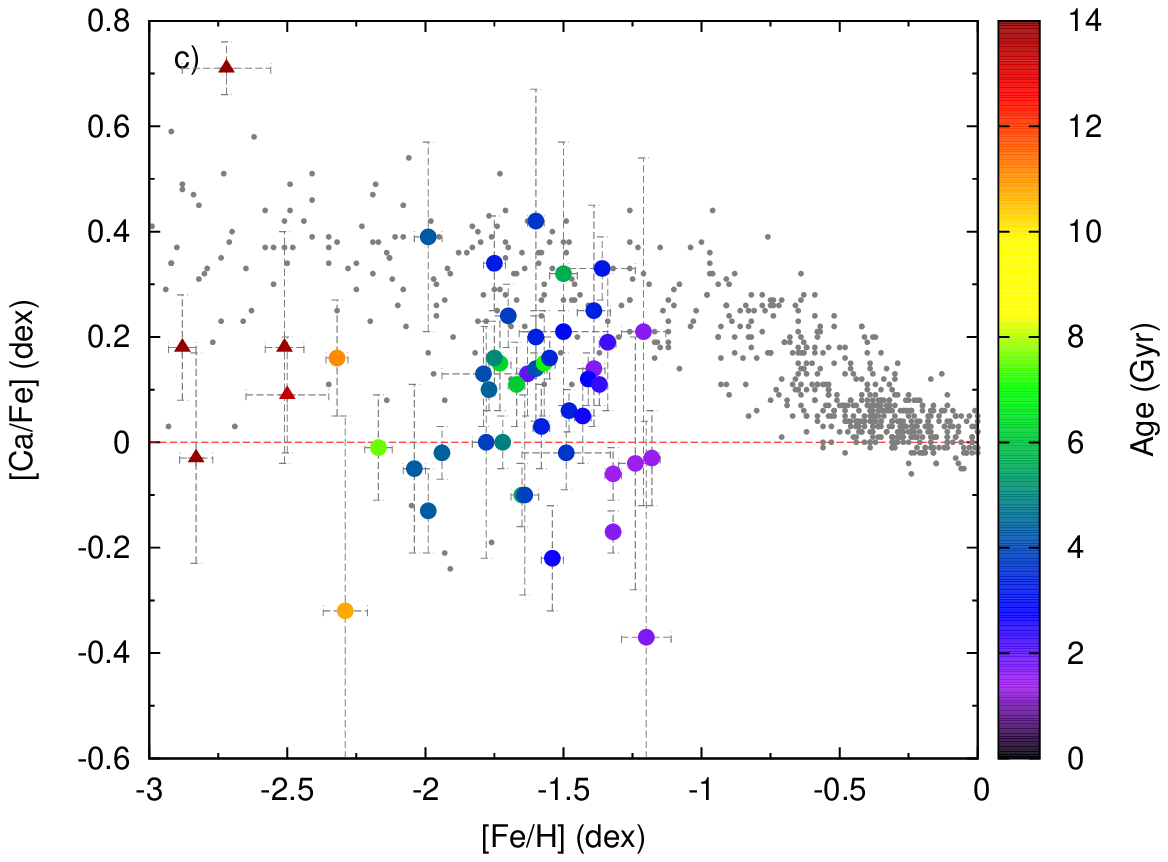}
\includegraphics[angle=0, width=0.35\textwidth]{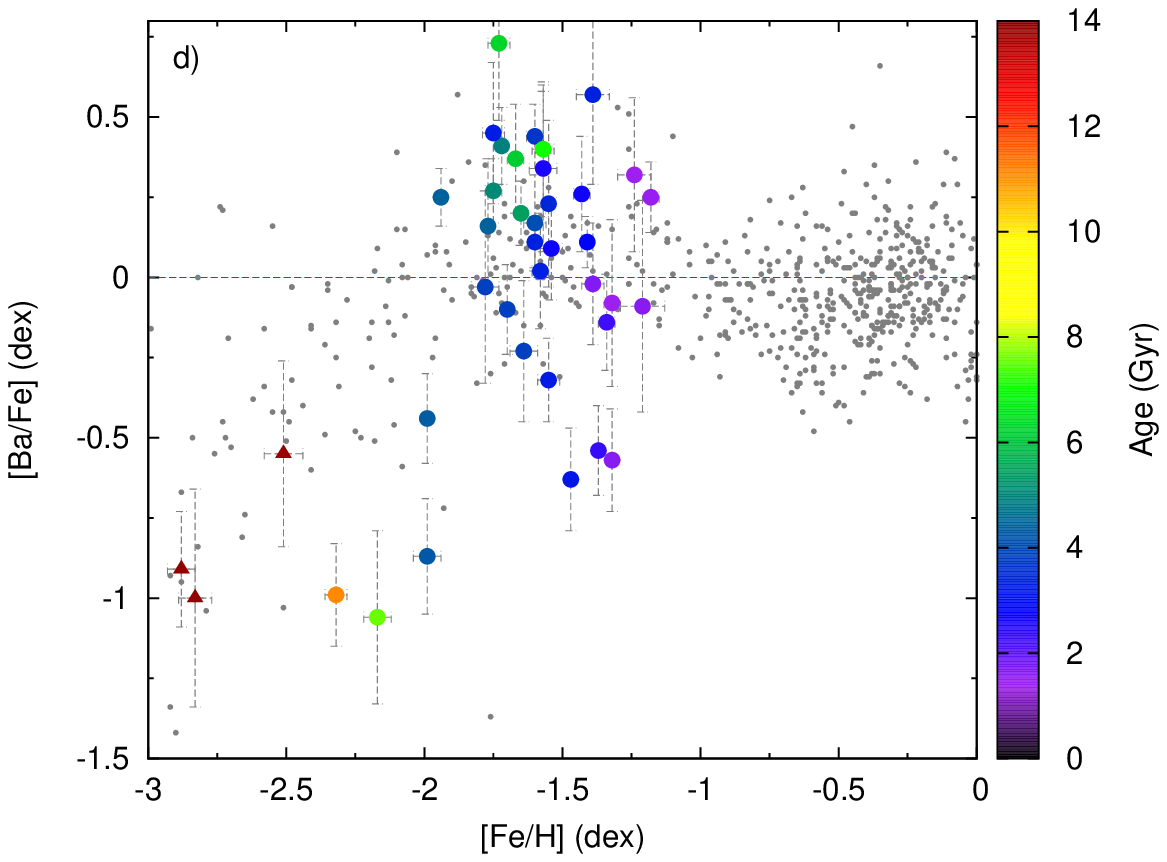}
\caption{Spectroscopic abundances of individual stars in the Carina dSph, with ages derived from the SFH determined using the Teramo isochrone set. The top-left panel shows the age-metallicity relation for individual stars from \ion{Ca}{ii} triplet spectroscopy~(red) and HR~(blue) observations. The abundances of [Mg/Fe]~(b), [Ca/Fe]~(c) and [Ba/Fe]~(d) are also shown, with colours indicating the age of individual stars. Triangles indicate stars for which no statistical age estimate could be determined. For those stars, only a rough age is given, based on the closest distance from evolutionary features in the SFH.  \label{Car123ages_TER}} 
\end{figure*}

The SFH solutions presented in Section~\ref{CarSFH} have been derived by adopting the Dartmouth isochrone set. To determine the effect of a change in isochrones on the SFH solution, we have also derived the SFH of Carina using the Teramo/BaSTI isochrone set~\citep{TeramoI}. The best-fitting SFH results for the Teramo isochrones are shown in Figures~\ref{Car123SFR_TER} and~\ref{Car123SFH_TER}, and can be directly compared to Figures~\ref{Car123SFR} and~\ref{Car123SFH}. The star formation rates as a function of age and metallicity derived from the SFH solution  using the BaSTI isochrones are given online  in Tables~2 and~3. 
\\
Figure~\ref{Car123residual_TER} shows the model CMD in comparison to the observed CMD for each annulus of Carina. The SFH in Figure~\ref{Car123SFH_TER} is roughly consistent with the solution derived using the Dartmouth isochrone. The age derived using the Teramo isochrone are are systematically shifted toward older ages, which is a well known effect caused by the difference in adopted colour transformations between both sets of isochrones. However, in both solutions, two main episodes of star formation are visible, showing a similar extent in age and metallicity. Therefore, our choice of isochrone set does not lead to considerable differences in the interpretation of the episodic SFH of Carina.  
\\
Using the best-fitting SFH results, we also derive the probability distribution for age for individual stars from spectroscopic observations. Stars with spectroscopic metallicities [Fe/H]$\le$$-$2.5 are treated as having [Fe/H]=$-$2.5 when determining ages, since BaSTI isochrones only extends down to this metallicity. The spectroscopic abundances of \ion{Ca}{ii} triplet and HR samples are shown in Figure~\ref{Car123ages_TER}, derived using the Teramo isochrones. Comparison to Figure~\ref{Car123ages} shows that very similar trends as a function of age are recovered, although the ages determined using the Teramo isochrones are consistently older by approximately 1 Gyr, as a result of the age difference in the SFH. 
\vspace{-1cm}

\end{appendix}

\end{document}